\documentclass[sigplan,nonacm,screen]{acmart}
\usepackage{xspace}
\usepackage{xurl}
\usepackage{hyperref}
\usepackage{url}
\usepackage{fancyhdr}
\usepackage[font=small,labelfont={small,bf}]{caption}
\usepackage[font=small,labelfont={small,bf}]{subcaption}

\newif\ifshowcomments
\showcommentsfalse 

\ifshowcomments
  \newcommand{\yang}[1]{\noindent{\color{magenta} {\bf \fbox{Yang}} #1}}
  \newcommand{\done}[1]{\noindent{\color{blue} {\bf \fbox{Done}} #1}}
  \newcommand{\lam}[1]{\noindent{\color{orange} {\bf \fbox{Lam}} #1}}
  \newcommand{\ziming}[1]{\noindent{\color{teal} {\bf \fbox{Ziming}} #1}}
\else
  \newcommand{\yang}[1]{}
  \newcommand{\done}[1]{}
  \newcommand{\lam}[1]{}
  \newcommand{\ziming}[1]{}
\fi

\newcommand{\parhead}[1]{\noindent\textbf{#1}}

\newcommand{\systemname}{UCCL-Zip\xspace}
\newcommand{\sys}{\textsc{\systemname}\xspace}
\newcommand{\sysp}{\textsc{Uzip-P2P}\xspace}
\newcommand{\sysnccl}{\textsc{Uzip-NCCL}\xspace}
\begin{document}
\fancyhead{}

\title[]{\sys: Lossless Compression Supercharged \\GPU Communication}

\newcommand{\gap}[0]{\hspace{0.2in}}
\author{
Shuang Ma$^{\dagger}$\gap
ChonLam Lao$^{\ddagger}$\gap
Zhiying Xu$^{\S}$\footnotemark[1]\gap
Zhuang Wang$^{\S}$\footnotemark[1]\gap
Ziming Mao$^{\P}$ \\
Delong Meng$^{\S}$\footnotemark[1]\gap
Zhen Jia$^{\S}$\footnotemark[1]\gap
Jun Wu$^{\S}$\footnotemark[1]\gap
Ion Stoica$^{\P}$\gap
Yida Wang$^{\S}$\footnotemark[1]\gap
Yang Zhou$^{\dagger}$ \\
\vspace{0.05in}
\textit{$^\dagger$UC Davis\gap
$^\ddagger$Harvard University\gap
$^\S$Amazon Web Services\gap
$^\P$UC Berkeley}
}

\begin{abstract}

The rapid growth of large language models (LLMs) has made GPU communication a critical bottleneck. While prior work reduces communication volume via quantization or lossy compression, these approaches introduce numerical errors that can degrade convergence, accuracy, and stability. 
We present \sys{}, a unified design that integrates lossless compression directly into GPU communication primitives. \sys{} supports both point-to-point (P2P) and collective communication without modifying user-facing APIs or compromising numerical correctness. For P2P communication, \sysp{} employs a split-send pipeline that exposes transmissible data early and overlaps compression with communication, while preserving high GPU efficiency by operating on large data blocks. 
For collective communication, \sysnccl{} integrates compression into NCCL’s persistent kernel model via fused execution, eliminating redundant memory traffic and kernel launches. In real workloads, \sys{} accelerates RL weight synchronization by up to 47.5\% and reduces vLLM end-to-end inference latency by up to 10\%, all without application changes.

\end{abstract}

\maketitle
\sloppy
\renewcommand{\thefootnote}{\fnsymbol{footnote}}
\footnotetext[1]{This work does not relate to the position at Amazon.}
\renewcommand{\thefootnote}{\arabic{footnote}}

\section{Introduction}

The rapid growth of model size and context length in large language models (LLMs)~\cite{google-palm,grattafiori2024llama3,deepseekai2025deepseekv3technicalreport,qwen2025qwen25technicalreport} has made multi-node, multi-GPU execution indispensable for both training and inference. In such settings, large volumes of data, including gradients, model parameters, activations, and intermediate states, are exchanged across GPUs within and across nodes. Collective communication~\cite{nccl,rccl,msccl} and point-to-point (P2P) communication are two widely used communication primitives in distributed GPU systems. Collective communication is commonly used in regular and structured communication patterns, such as tensor parallelism (TP), pipeline parallelism (PP), and other distributed training paradigms~\cite{megatron-lm,deepspeed,colossalai}. P2P communication provides greater flexibility and is commonly used in scenarios such as KV cache transfer~\cite{uccl_p2p,nixl2025,pplx2025,mooncake_te} and asynchronous weight updates in reinforcement learning workloads. As a result, GPU communication performance has become a critical bottleneck~\cite{crux,cassini,chang2024flux}, directly impacting the efficiency and scalability of LLM training and inference.

To reduce communication overhead, prior work has extensively explored quantization and other lossy compression techniques to reduce data volume~\cite{qsgd2017,ZeRO2023,THC2024,gzcccl2024,grace2021,compso2025}. While effective in improving bandwidth utilization, these approaches inevitably introduce numerical errors. These errors can slow convergence during training or degrade model accuracy at inference time, and may also cause training–inference mismatch in RL~\cite{he2025nondeterminism}. More recently, lossless compression techniques on GPUs have begun to emerge~\cite{dietgpu_github,ZipNN2025,NeuZip2024,dfloat11,nvcomp2023,azami2025lossless}, demonstrating the potential to reduce data size without sacrificing numerical fidelity. This observation opens up a promising opportunity: applying lossless compression to GPU communication. However, despite its potential, this direction remains largely underexplored, particularly in the context of tightly integrated communication pipelines.

\begin{figure}[t!]
\centering

\begin{subfigure}[t]{0.95\columnwidth}
    \centering
    \includegraphics[width=\linewidth]{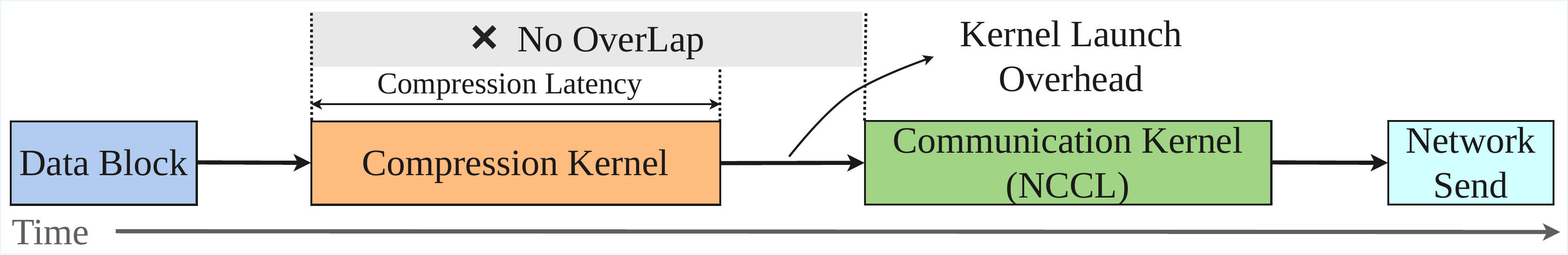}
    \caption{Baseline.}
    \label{fig:overview_a}
\end{subfigure}


\begin{subfigure}[t]{0.48\columnwidth}
    \centering
    \includegraphics[width=\linewidth]{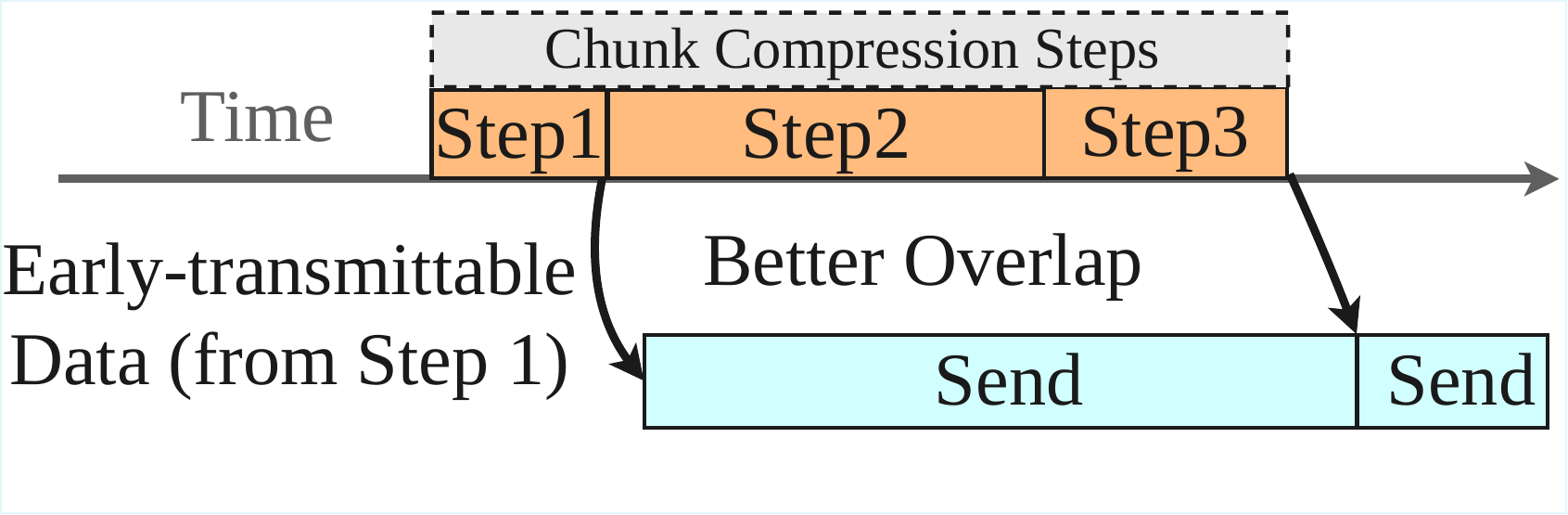}
    \caption{\sysp{}.}
    \label{fig:overview_b}
\end{subfigure}
\hfill
\begin{subfigure}[t]{0.48\columnwidth}
    \centering
    \includegraphics[width=\linewidth]{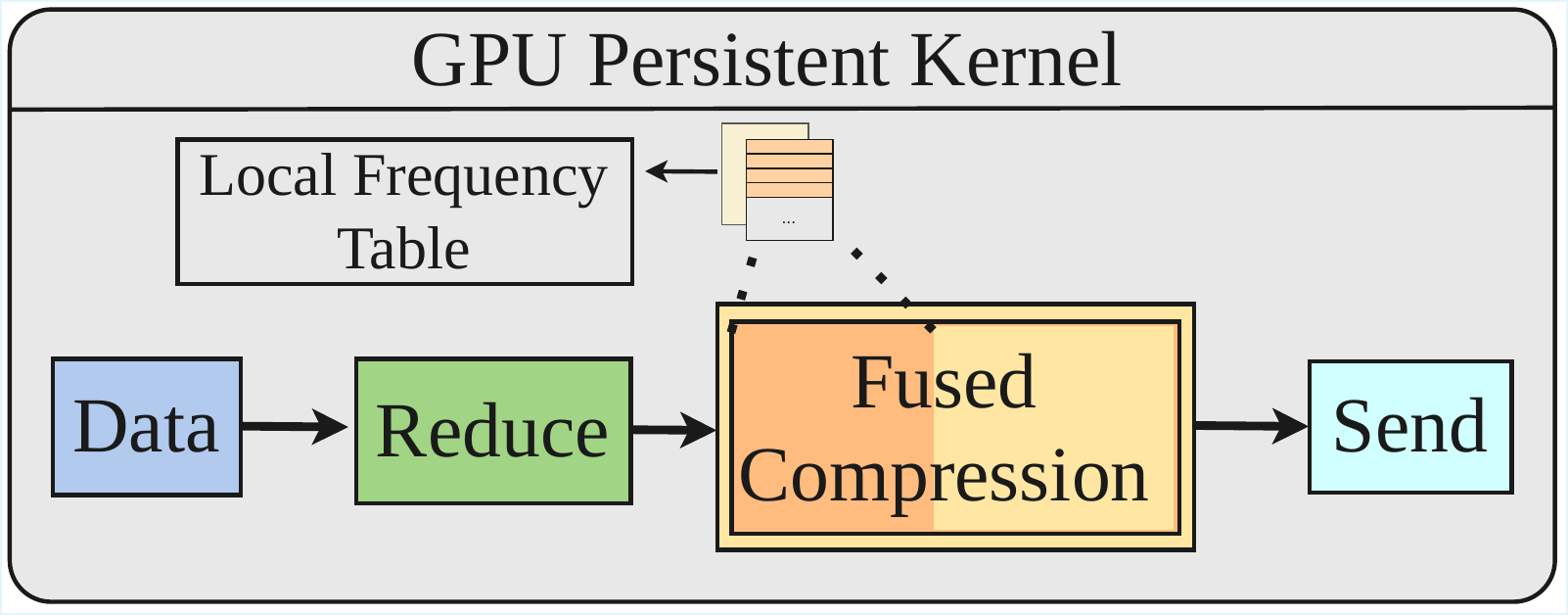}
    \caption{\sysnccl.}
    \label{fig:overview_c}
\end{subfigure}

\caption{
Overview of \sys{}. 
(a) Naive design suffers from lack of overlap and additional kernel overhead.
(b) \sysp{} enables early transmission and overlaps compression with communication; detailed steps are shown in Figure~\ref{fig:p2p_pipeline}.
(c) \sysnccl{} integrates compression into the persistent kernel, eliminating redundant memory traffic and extra kernel launches.
}

\label{fig:overview}

\end{figure}

Effectively hiding the overhead of lossless compression within the communication pipeline remains a key challenge. On modern GPUs, compression latency does not scale linearly with input size due to their highly parallel execution model, where throughput is maximized only when sufficient parallelism is exposed to fully utilize hardware resources~\cite{nvidiaCUDAprogramming}. As a result, compressing smaller chunks often incurs similar latency to larger ones, undermining conventional fine-grained pipelining strategies that partition data into small blocks and overlap compression with transmission. In such designs, excessive fragmentation can even amplify overheads, leading to degraded end-to-end performance.

As shown in Figure~\ref{fig:overview_a}, this challenge is further exacerbated in collective communication. Libraries such as NCCL launch a single kernel per message to orchestrate computation and communication. Data is processed at a fine granularity, typically on the order of a few kilobytes per chunk, and staged through internal buffers to enable overlap. 
Directly inserting a standalone compression stage into this workflow introduces additional kernel launches and redundant memory traffic, which can negate the benefits of compression.

In this work, we present \sys{}, a unified design that integrates lossless compression directly into GPU communication primitives, thereby addressing the fundamental tension between compression overhead and communication efficiency. \sys{} seamlessly supports both collective and P2P communication, enabling bandwidth reduction without modifying user-facing APIs or compromising numerical correctness. Our approach consistently improves effective bandwidth and, in many cases, approaches or exceeds the hardware-imposed network limit on modern GPU clusters.

For P2P communication, \sysp{} employs a compression-aware \textit{split-send} pipeline that aligns with the structure of floating-point compression (Section~\ref{subsec:design_p2p}). As shown in Figure~\ref{fig:overview_b}, \sysp{} processes data in sufficiently large chunks to maintain high GPU utilization, and decouples compression stages to expose data that can be transmitted early. This allows a significant portion of the payload to be sent over the network before full compression is complete, effectively overlapping communication with compression.

For collective communication, such as AllReduce, \sysnccl{} integrates compression into NCCL’s persistent kernel execution model (Section \ref{subsec:design_Uzip_nccl}). As shown in Figure~\ref{fig:overview_c}, we redesign the compression workflow by introducing localized frequency tables that enable block-wise, on-the-fly entropy modeling, eliminating the need for global preprocessing. We then fully fuse compression into the compute–communication pipeline, avoiding extra kernel launches and reducing redundant HBM accesses.

By aligning compression with the execution granularity of the collective loop, compression and communication proceed in a tightly coupled streaming fashion without introducing additional synchronization or kernel launches.

\sys{} is designed to be portable across heterogeneous hardware platforms. It supports multiple GPU architectures, including AMD and NVIDIA devices, and operates over diverse RDMA-capable interconnects such as NVIDIA CX7, Broadcom Thor-2, and AWS EFA NICs. In addition, \sys{} extends DietGPU~\cite{dietgpu_github} to support float8 compression, enabling compatibility with a wide range of floating-point formats commonly used in modern machine learning workloads, including \texttt{bfloat16}, \texttt{float16}, \texttt{float32}, \texttt{float8\_e4m3fn}, and \texttt{float8\_e5m2}. 
This broad compatibility enables \sys{} to be readily deployed in existing training and inference systems without requiring changes to data representation or communication backends.

We evaluate \sys{} across a diverse set of GPU clusters, including multi-node RDMA environments with EFA and RoCEv2 (Section~\ref{sec:eval}).  In reinforcement learning training, \sysp improves weight synchronization throughput by up to 47.5\% on large tensors. In distributed LLM inference with vLLM, \sysnccl reduces end-to-end inference latency by up to 10\%. These benefits are achieved transparently, without modifying application code. 
We also reveal the architecture incompatibility of the widely-used NCCL with lossless compression, and identify concrete future directions in designing a better collective library (e.g., two-shot collectives and increasing chunk sizes).
\sys has been open-sourced at https://github.com/uccl-project/uccl/tree/main/p2p.

\section{Background and Motivation}

\subsection{Basic GPU Lossless Compression Design}
\label{dietgpu_background}
\begin{table}[t]
\centering
\setlength{\tabcolsep}{4pt}
\small
\begin{tabular}{lccc}
\toprule
\textbf{Tensor Type} & \textbf{Size (MB)} & \textbf{Data Type} & \textbf{Compression Ratio} \\
\midrule
Gradient   & 214  & FP32  & 0.848 \\
Activation & 16   & BF16  & 0.679 \\
Weight     & 107  & BF16  & 0.675 \\
\bottomrule
\end{tabular}
\caption{Compression ratios of representative tensors collected during GLM4-9B training~\cite{glm45v2025}. 
We define the compression ratio as the ratio between the compressed size and the original (uncompressed) size; thus, lower values indicate better compression. }
\label{tab:compression_ratio}
\end{table}

\subsubsection{Statistical Structure of Floating-Point Values}
Many existing works~\cite{dietgpu_github,dfloat11,NeuZip2024,ZipNN2025} exploit the statistical properties of floating-point tensors to enable efficient lossless compression. 
Floating-point values consist of three fields: sign, exponent, and fraction (mantissa). 
Among these components, the fraction and sign bits typically exhibit near-uniform distributions 
and thus provide limited compression opportunities. In contrast, the exponent field often 
follows a narrow and highly skewed distribution in many machine learning workloads. This skew becomes more pronounced when tensors are normalized to a bounded range.

Consider the \texttt{bf16} format, which contains 1 sign bit, 8 exponent bits, and 7 fraction bits. 
When a bf16 tensor is normalized, the exponent field frequently collapses to a small 
number of repeated patterns. 
Such highly skewed symbol distributions are well suited for entropy coding techniques such as Asymmetric Numeral Systems (ANS)~\cite{nvcomp2023,dietgpu_github,azami2025lossless}, enabling high compression ratios. 
Table~\ref{tab:compression_ratio} reports compression ratios of representative tensors from GLM4-9B~\cite{glm45v2025} training. The ratio remains stable across tensors of the same type: \texttt{float32} gradients achieve 0.85, while \texttt{bfloat16} tensors consistently achieve around 0.68 across layers and parameter types.

\subsubsection{GPU Compression Algorithm for Floating-Point Tensors}

Prior work, such as DietGPU~\cite{dietgpu_github}, leverages GPUs to accelerate compression of floating-point tensors. 
Exploiting the statistical structure described above, such approaches decompose each value into two components: the exponent field and the remaining bits (sign and fraction), and compress only the exponent field.
This design improves compression efficiency compared to directly compressing raw floating-point representations.
Figure 2 illustrates the overall structure of a typical GPU-based ANS compression pipeline, which consists of three steps, each performing a single pass over global memory.

\begin{figure}[!t]
\centering
\includegraphics[width=0.95\columnwidth]{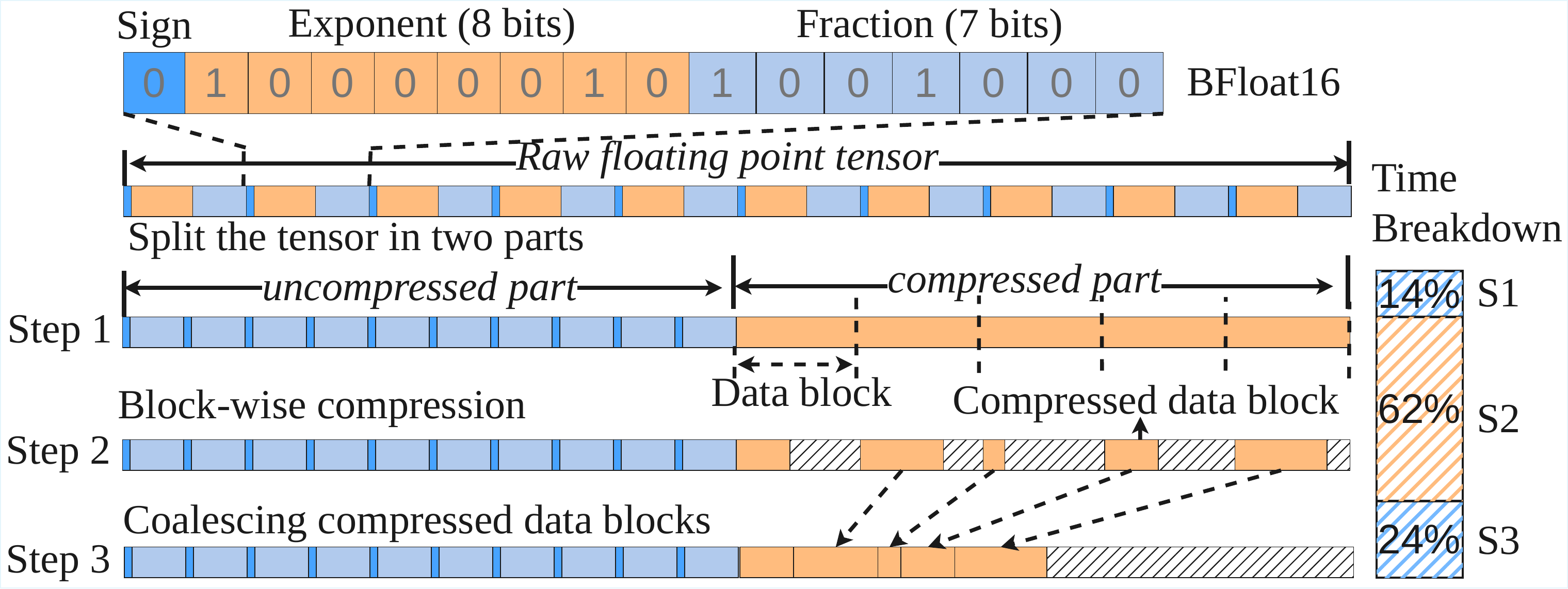}
\caption{Overview and time breakdown of a typical GPU-based floating-point ANS compression pipeline (S1–S3 denote Steps 1–3). }
\label{fig:compression_pipeline}
\end{figure}

\parhead{Step 1: Float Splitting and Frequency Table Construction.} During the first pass, each value is decomposed into its exponent and remaining bits (sign and fraction) (Figure~\ref{fig:compression_pipeline}), which are written to two contiguous buffers in global memory: the \textit{compressed part} (exponents) and the \textit{uncompressed part}. Then each 8 bit segment in the compressed part is interpreted as a symbol. During the split step, the kernel simultaneously collects symbol frequencies, constructing the \emph{global frequency table} required by the ANS encoder.

\parhead{Step 2: Independent Block-wise Compression.}
Once the frequency table is constructed, multiple GPU thread blocks perform compression 
independently using the shared global frequency table (Figure~\ref{fig:compression_pipeline}). 
Each thread block reads a chunk of exponent symbols from global memory and applies 
ANS encoding locally. Entropy coding produces variable-length outputs. Therefore, each block writes compressed data to a temporary global memory buffer instead of the final output buffer.

\parhead{Step 3: Coalescing Compressed Blocks.} In the final step, the variable-length outputs generated by different blocks are merged 
into a single contiguous output buffer (Figure~\ref{fig:compression_pipeline}). 
The coalescing process introduces a third global memory write to assemble the final compressed stream.

\subsection{NCCL Collectives Implementation}

\label{nccl_background}
Modern GPU collective libraries such as NCCL~\cite{nccl} adopt a channel-based execution model, where the input tensor is partitioned into multiple chunks and processed by independent channels in parallel. Each channel is typically mapped to a Cooperative Thread Array (CTA), which is responsible for moving and reducing a disjoint portion of the data through the communication topology (e.g., ring or tree)~\cite{hu2025demystifying,msccl}. 

To maximize overlap between communication and computation, NCCL decomposes communication into fine-grained slices that are pipelined across CTAs, with each CTA operating largely independently on its assigned data range. This design avoids global synchronization and enables high hardware utilization, but also limits coordination across CTAs within a kernel. In particular, CTAs do not cooperatively share intermediate statistics or synchronize frequently during execution, as such coordination would break the pipeline parallelism and incur substantial overhead.

As a result, operations requiring global state across the entire tensor (e.g., constructing a global frequency table for ANS encoding) are fundamentally incompatible with NCCL's execution model. Supporting such operations would require either an additional kernel launch or explicit cross-CTA barriers, both of which break pipeline continuity and preclude tight integration with NCCL's collective execution.

\begin{figure}[!t]
\centering
\begin{subfigure}[t]{0.22\textwidth}
    \centering
    \includegraphics[width=\textwidth]{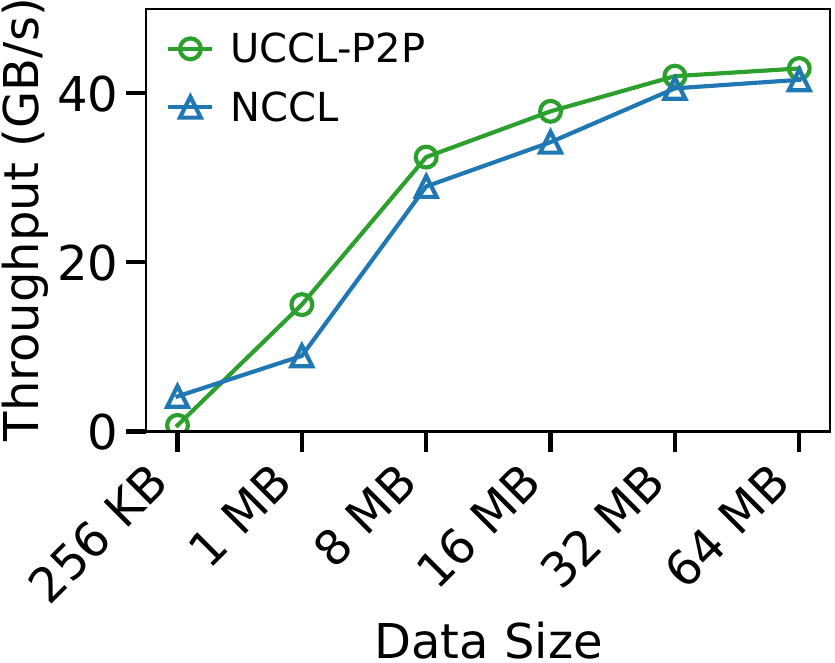}
    \caption{P2P throughput. 
    }
\end{subfigure}
\hfill
\begin{subfigure}[t]{0.22\textwidth}
    \centering
    \includegraphics[width=\textwidth]{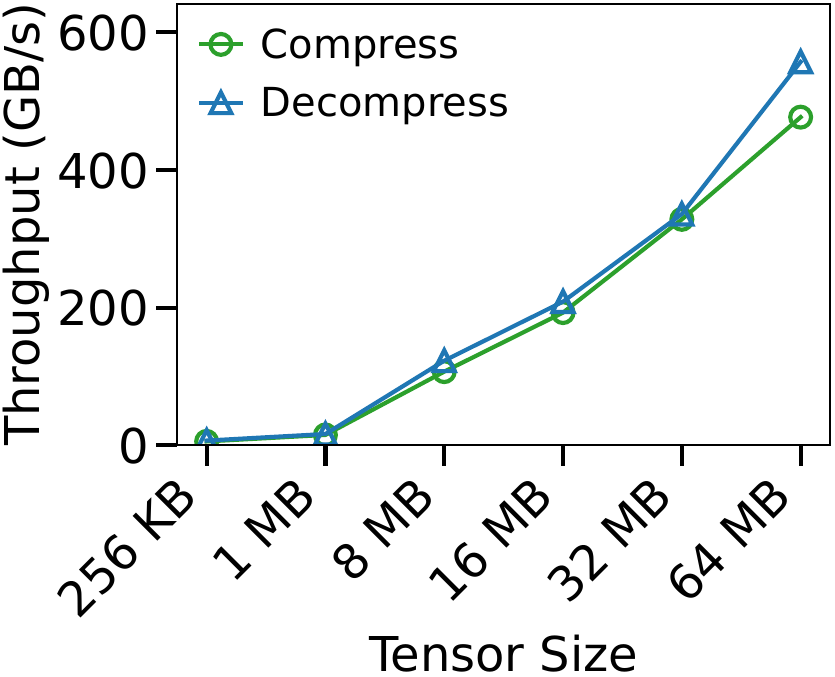}
    \caption{Compression throughput.}
\end{subfigure}

\caption{Throughput of P2P communication and DietGPU compression on AWS p5en.48xlarge (two nodes, EFA). 
}
\label{fig:p2p_dietgput}
\end{figure}

\subsection{Challenges of Applying Lossless Compression to GPU Communication}

\parhead{Challenge 1: Compression latency must not offset communication savings.} The latency of compression can outweigh the savings from reduced data transfer. As we show, conventional approaches to hiding compression latency often yield negligible gains or even degrade performance.

A naive design compresses all data prior to transmission, assuming reduced data volume will improve throughput. Consider a representative example in RL training pipelines, where model weights are transferred between training and rollout nodes via P2P send/recv. A typical tensor size is 16\,MB (e.g., bf16 tensors of shape $[4096, 2048]$ in MoE models such as Qwen3-35B-A22B~\cite{qwen}). As shown in Figure~\ref{fig:p2p_dietgput}, compressing a 16\,MB tensor incurs $\sim$90\,$\mu$s of latency, while the reduced data volume saves only $\sim$108\,$\mu$s in communication time, resulting in a marginal throughput improvement of $\sim$4\%.

A natural mitigation is to partition tensors into chunks and pipeline compression with communication. However, this strategy is ineffective for GPU-based compression. Splitting a 16\,MB tensor into four 4\,MB chunks does not alleviate the bottleneck. Each 4\,MB chunk still incurs $\sim$70\,$\mu$s of compression latency, while the per-chunk communication savings shrink to only $\sim$20\,$\mu$s, resulting in a net performance degradation. This asymmetry stems from GPU inner parallelism: compression latency does not scale linearly with data size, as the GPU remains highly unutilized for small inputs.

\parhead{Challenge 2: Compression must be co-designed with GPU communication frameworks.}
Modern frameworks such as NCCL rely on persistent kernels and chunk-based pipelining, which impose fundamental barriers to integrating compression. First, existing compression approaches introduce global synchronization, conflicting with the persistent kernel design. Second, chunk-based pipelines require minimal data movement to sustain throughput; naive on-GPU compression introduces additional global memory copies, significantly degrading communication efficiency. Taken together, existing lossless compression methods are impractical when naively integrated into GPU communication stacks.

\section{Design}
\label{sec:design}

\subsection{Overview}
\sys{} aims to address the above challenges when applying lossless compression to GPU communication. 

For P2P scenarios (Section \ref{subsec:design_p2p}), \sys{} introduces a \textit{split-send} pipeline that overlaps compression with network transmission. Unlike naive pipelining schemes, \sysp{} aligns pipeline stages with the internal steps of the compression algorithm. This stage-aligned design operates on large data blocks rather than fine-grained chunks, preserving GPU efficiency while exposing data that can be transmitted early in the compression process. Consequently, network transmission begins earlier and overlaps with subsequent compression steps, enabling faster bandwidth ramp-up and improved network utilization.

For NCCL scenarios (Section \ref{subsec:design_Uzip_nccl}), \sys{} integrates compression into the NCCL data path via a fused kernel (Section \ref{subsec:design_fused_kernel}). The kernel combines Steps~1 and~2 in Figure~\ref{fig:compression_pipeline} using a redesigned frequency table aligned with NCCL’s execution model, reducing global memory traffic. Step~3 is eliminated via warp-level execution, directly writing compressed outputs to a GPU-resident FIFO. 

\subsection{Split-Send Communication Pipeline}
\label{subsec:design_p2p}

To hide compression overhead in the send/recv path, we design a
\textit{split-send} communication pipeline that overlaps compression
with network transmission. Unlike conventional chunk-based pipelining that partitions tensors into independent blocks, our design aligns pipeline
boundaries with the internal steps of the compression algorithm.

\textit{Split-send} is motivated by the observation that the initial stage of the compression pipeline (Step~1 in Figure~\ref{fig:compression_pipeline}) incurs relatively low latency, while subsequent stages are more compute-intensive.
Once Step~1 completes, \sys{} immediately initiates transmission of the corresponding uncompressed data. 
This early transmission overlaps with the execution of later compression stages (Step~2 and Step~3), which dominate the overall compression cost.
After compression completes, only a small volume of compressed data remains to be transmitted.
By exposing transmissible data early and overlapping communication with compute-heavy compression stages, \textit{split-send} reduces communication stall time and improves end-to-end throughput.

\subsubsection{Key Properties of GPU Compression}

Our design is motivated by two key properties of GPU-based floating-point
compression.

\parhead{Property 1: Compression latency scales sub-linearly with data size.}
Due to massive GPU parallelism and kernel launch overheads, the latency of
GPU compression increases much more slowly than the input size. Consequently, compressing small chunks may take nearly as long as
compressing substantially larger ones. For example, on an NVIDIA H200 GPU, compressing a 16\,MB floating-point tensor takes approximately 90\,$\mu$s, while compressing a 4\,MB tensor takes approximately 70\,$\mu$s. Although the data size is reduced to 1/4, the latency decreases modestly by only 22\%. This behavior indicates that splitting tensors into many small chunks can significantly reduce GPU compression efficiency.

\begin{figure}[t]
\centering
\includegraphics[width=0.95\columnwidth]{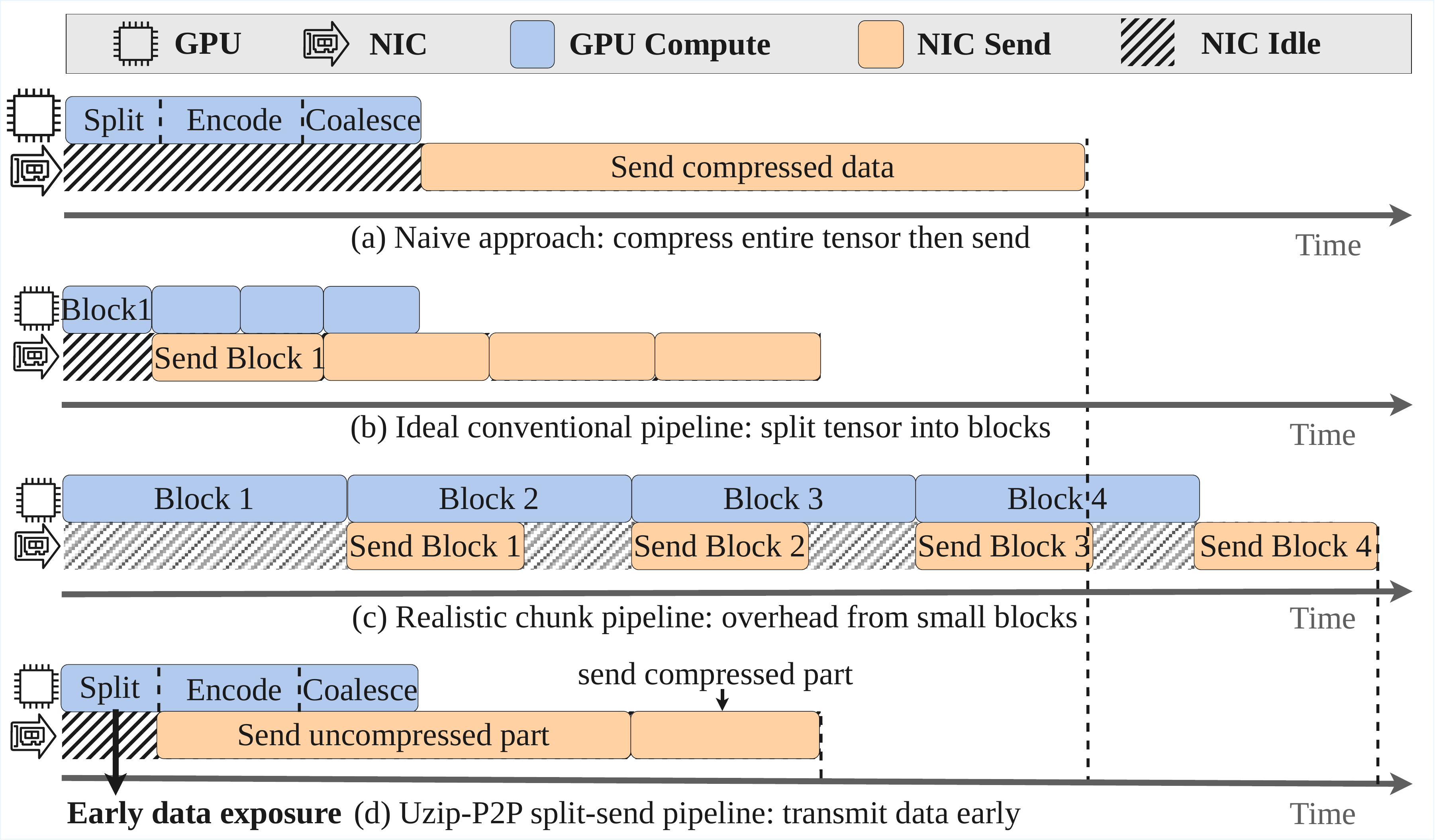}
\caption{Comparison of pipelining designs for overlapping compression and communication. (b)(c) Chunk-based pipelining assumes latency scales with size but is ineffective for GPU compression. (d) Split-send exposes transmissible data early and overlaps communication with remaining compression.}
\label{fig:p2p_pipeline}
\end{figure}

\parhead{Property 2: Early availability of transmissible data.}
The first step of the compression pipeline accounts
for 14\% of the total compression time but already exposes a
large portion of the final representation (Figure~\ref{fig:compression_pipeline}). The first stage of the compression pipeline performs a structural split of
the input tensor into two components: an \textit{uncompressed part} and a
\textit{compressed part}. The ratio between the two portions depends on the floating-point format.
For example, for \texttt{bf16} tensors, the uncompressed and compressed
portions each account for approximately half of the tensor. For
\texttt{float32}, the uncompressed portion occupies roughly three quarters
of the data. Since the compressed portion will be further reduced by the
subsequent compression stages, the uncompressed portion constitutes an
even larger fraction of the final transmitted data. Importantly, once the split stage completes, the uncompressed
portion requires no further computation and can therefore be transmitted
immediately. 

\subsubsection{Limitations of Chunk-Based Pipelines}

Conventional pipelining splits tensors into independent chunks to reduce
the latency of producing the first transmittable byte and to overlap
computation with communication. As shown in Figure~\ref{fig:p2p_pipeline} (b), this approach assumes that compression
latency decreases proportionally with data size.

However, this assumption does not hold for GPU compression. Compression
involves multiple GPU kernels and memory transformations whose overheads
do not scale with input size. Consequently, compressing a small chunk
may incur nearly the same latency as compressing a large tensor while
significantly reducing overall GPU throughput. As shown in Figure~\ref{fig:p2p_pipeline} (c), naïve chunking can therefore
degrade performance in compression–communication pipelines.

\subsubsection{Split-Send Pipeline Design}

Based on these observations, we design a \textit{split-send communication pipeline} that overlaps compression with communication while preserving
large-block GPU efficiency.

After the first compression stage completes, we divide the
tensor into two components:

\begin{itemize}
\item \textbf{Uncompressed portion.} This portion is finalized after the
first stage and can be transmitted immediately.
\item \textbf{Compressed portion.} The remaining data continues through
the subsequent compression stages before transmission.
\end{itemize}

As shown in Figure~\ref{fig:p2p_pipeline} (d), while the GPU processes the remaining compression stages, the network interface concurrently transmits the uncompressed portion.
Once compression finishes, the smaller compressed payload is transmitted.
This pipeline exposes data early and o`verlaps compression with network
communication without sacrificing GPU throughput.

\subsection{Fused Compression Kernel}

\label{subsec:design_fused_kernel}
To integrate compression into NCCL efficiently, we fuse the multi-step compression pipeline (in Figure~\ref{fig:compression_pipeline}) into a single fused kernel execution. 
The first two steps are merged through localized frequency tables, while the third coalescing step is eliminated by directly streaming compressed outputs into the communication pipeline.
This design simplifies integration with NCCL primitives and improves performance by reducing global memory traffic and avoiding extra kernel launch overhead.

\begin{figure}[t!]
\centering
\begin{subfigure}[t]{0.45\columnwidth}
    \centering
    \includegraphics[width=\linewidth]{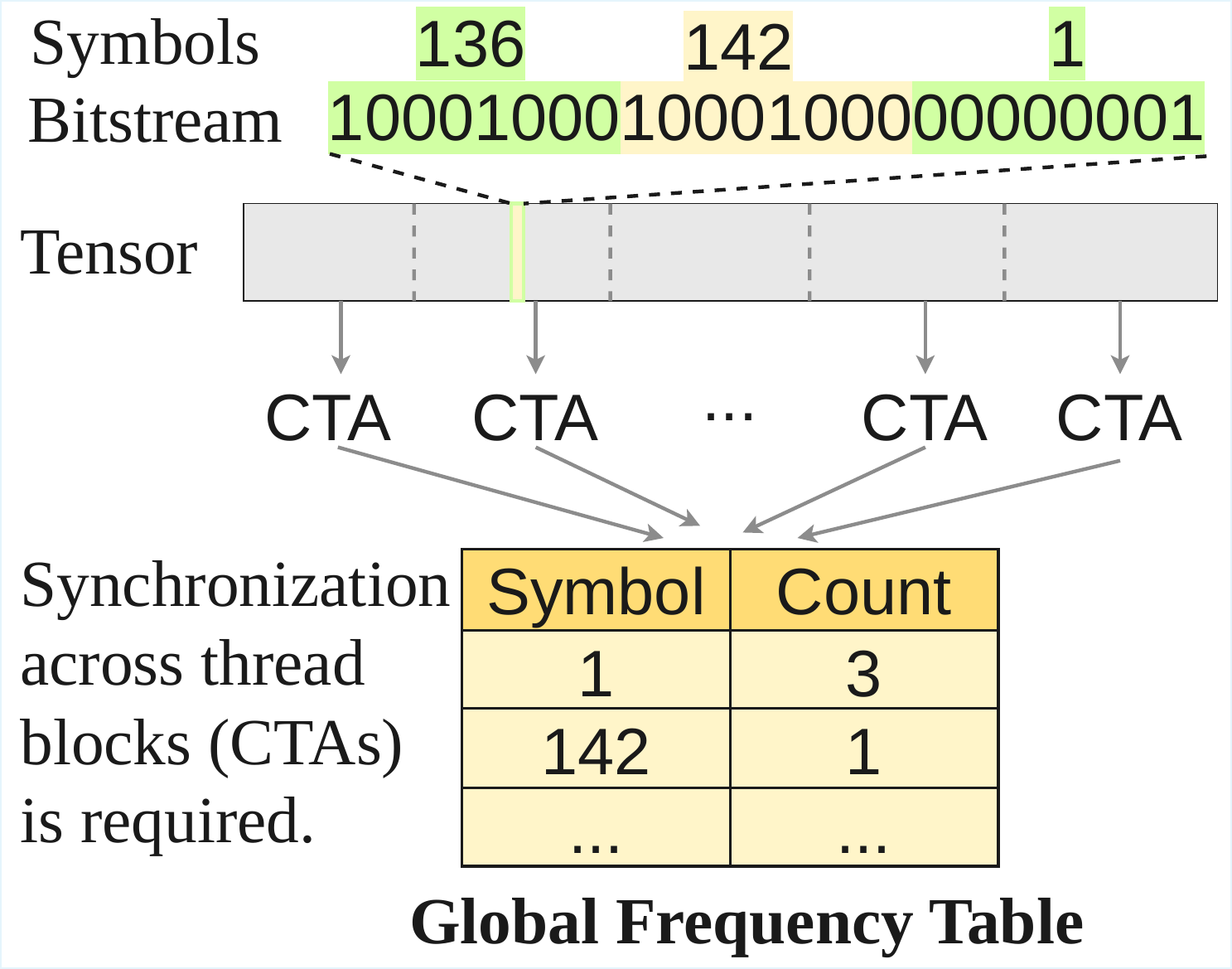}
    \caption{Global frequency table construction.}
    \label{fig:local_frequency_table_a}
\end{subfigure}
\hfill
\begin{subfigure}[t]{0.45\columnwidth}
    \centering
    \includegraphics[width=\linewidth]{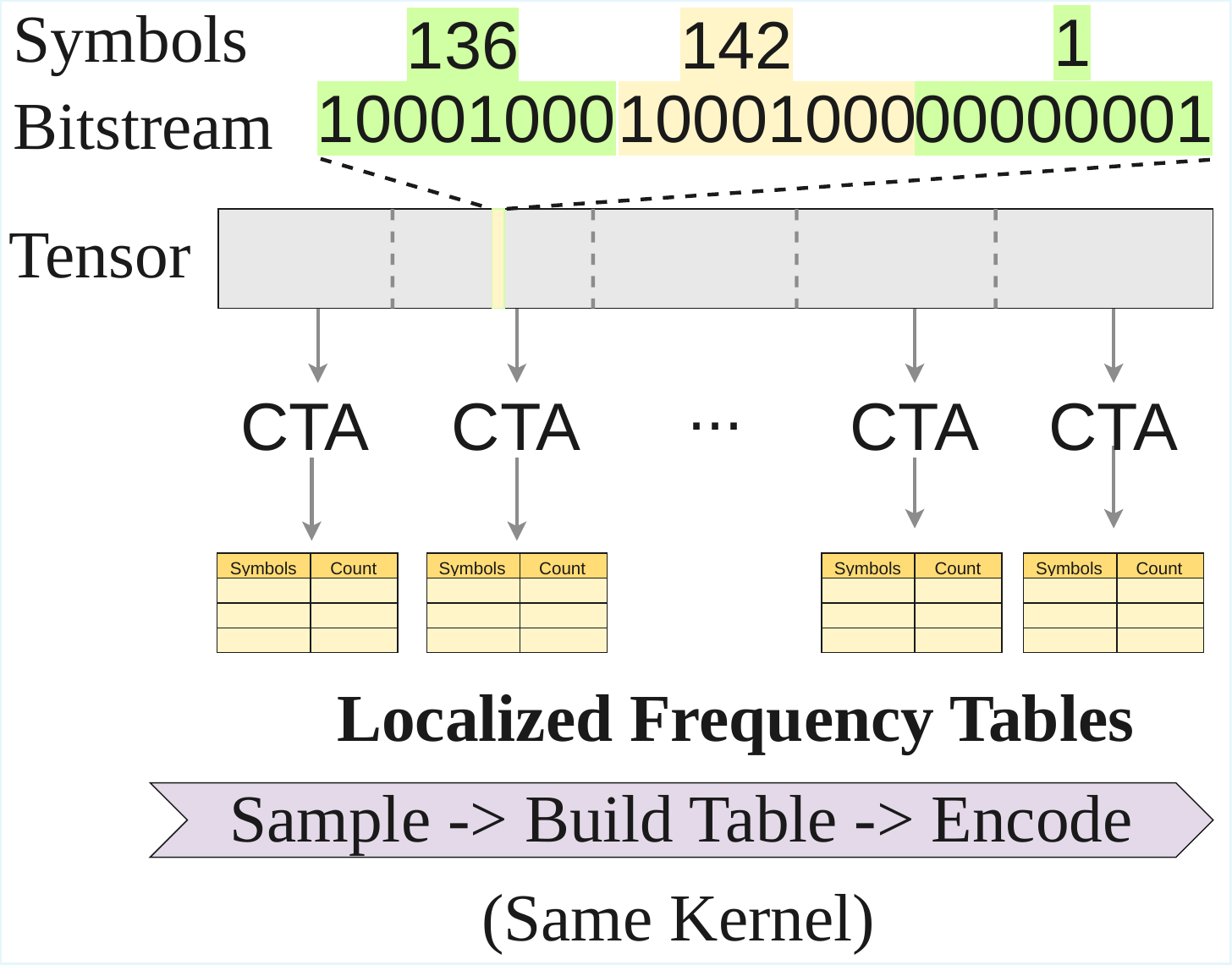}
    \caption{Localized frequency tables.}
    \label{fig:local_frequency_table_b}
\end{subfigure}

\begin{subfigure}[t]{0.48\columnwidth}
    \centering
    \includegraphics[width=\linewidth]{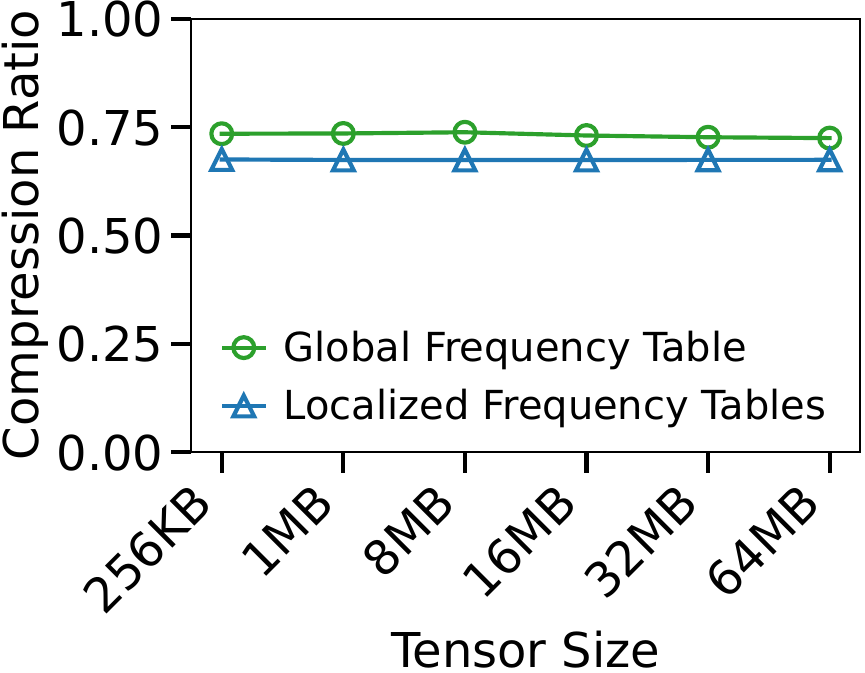}
    \caption{Compression ratio.}
    \label{fig:local_frequency_table_c}
\end{subfigure}

\caption{
Localized frequency tables eliminate global coordination and enable a fully fused compression pipeline. 
(a) A global frequency table requires cross-CTA synchronization, preventing kernel fusion and introducing additional memory passes. 
(b) With localized tables, each CTA independently samples and constructs its own table, enabling a fused pipeline within a single kernel without synchronization. 
(c) This design preserves compression effectiveness.
}
\label{fig:local_frequency_table}
\end{figure}

\subsubsection{Localized Frequency Tables}
The original design (Section \ref{dietgpu_background}), such as in DietGPU, 
builds a global frequency table by scanning the entire tensor (Figure~\ref{fig:local_frequency_table_a}), which is incompatible with NCCL's channel-based execution model. This is because NCCL CTAs process disjoint data partitions independently without synchronization, preventing efficient global coordination.
As a result, global frequency table construction prevents fusing the first two stages in Figure~\ref{fig:compression_pipeline} into a single kernel and incurs an additional round of memory access. 

We design localized frequency tables (Figure~\ref{fig:local_frequency_table_b}) to enable kernel fusion with a comparable compression ratio and lower synchronization overhead. 

Instead of maintaining a shared global frequency table, each thread block constructs its own local table by sampling a small portion of its assigned data range (e.g., the first 256\,KB). 
This localized sampling captures the exponent distribution without a full traversal, enabling fusion into a single kernel and reducing global memory accesses to one pass.

Local frequency estimation preserves compression effectiveness while eliminating synchronization overhead. Because exponent distributions in neural network tensors exhibit stable statistical structure and are concentrated~\cite{gauss}, tables constructed from sampled subsets of each block approximate the global distribution well, achieving compression ratios close to the global-table baseline while removing an entire memory pass.
This introduces only a minor approximation: it may not fully capture rare exponent values, but still preserves near-optimal coding efficiency. As shown in Figure~\ref{fig:local_frequency_table_c}, on real tensor data, this design incurs only about a 4.5\% reduction in compression ratio, and this small cost remains consistent across different tensor sizes.
Each block then uses its local table for both compression and decompression, eliminating the need for global frequency table lookups or cross-block synchronization. 

\vspace{-5pt}
\subsubsection{Eliminating the Third Coalescing Step.}
We further optimize the pipeline by having each thread block write compressed outputs directly into NCCL’s FIFO buffer, eliminating the third coalescing step in the original workflow (Figure~\ref{fig:compression_pipeline}).
NCCL internally maintains a staging buffer, where user data is first copied before being transmitted over the network. In our design, we directly write the compressed outputs produced in Step 2 (Figure~\ref{fig:compression_pipeline}) into this NCCL-managed buffer.
This enables a fully fused single-kernel implementation, reducing global memory accesses from three passes to one and avoiding multiple kernel launches.

\subsection{Compression-Integrated NCCL Pipeline}
\label{subsec:design_Uzip_nccl}

We integrate the fused compression kernel directly into NCCL’s
collective communication pipeline, eliminating extra launches and operating directly on data in registers or shared memory. This design allows compression,
communication, and decompression to execute in a tightly overlapped
pipeline without requiring any changes from user applications.

\begin{figure}[t]
\centering
\includegraphics[width=0.95 \columnwidth]{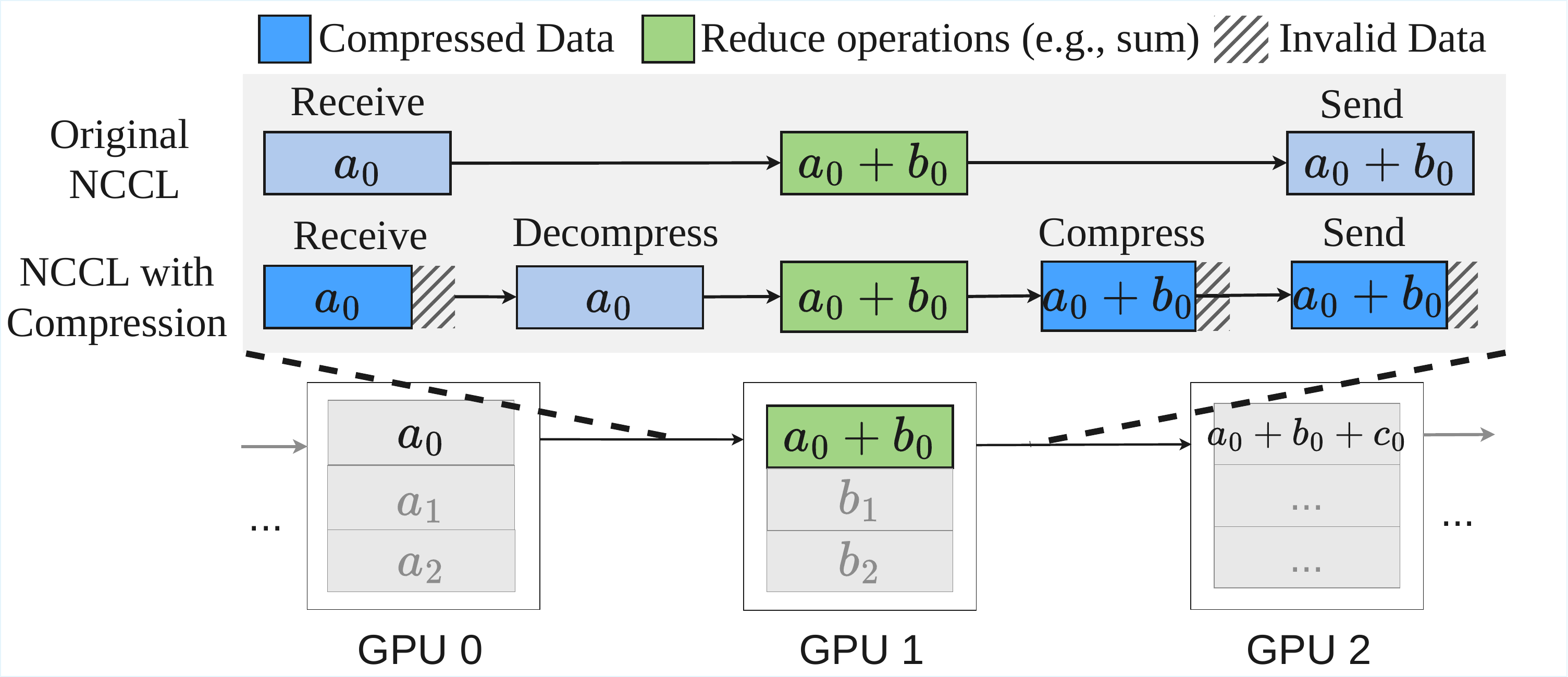}
\caption{Overview of NCCL \texttt{all\_reduce} with compression.
The original NCCL pipeline performs \textit{receive–reduce–send} operations per slice. We extend it by integrating decompression and compression into the dataflow. 
}
\label{fig:nccl_overview}
\end{figure}

Figure~\ref{fig:nccl_overview} illustrates the workflow of NCCL \texttt{all\_reduce} with compression.
In the original NCCL pipeline, each GPU performs a sequence of
\textit{receive–reduce–send} steps for every data slice.
Our design extends this pipeline by inserting compression and
decompression stages while preserving the streaming execution model. Specifically, data is compressed at the sender, transmitted in compressed form, and decompressed at the receiver before reduction. The reduced result is then recompressed and forwarded to the next GPU
in the ring. For example, GPU~1 receives compressed slice $a_0$ from GPU~0, decompresses it, reduces it with local slice $b_0$ to produce $a_0 + b_0$, and then recompresses and forwards the result to the next GPU 2.

\parhead{Integration into the NCCL data path.}
We integrate compression directly into NCCL’s data processing layer,
where collective primitives such as \texttt{all\_reduce}, \texttt{all\_to\_all},
and \texttt{send/recv} are implemented.
The key modification occurs in the stage previously handled by
\texttt{CopyReducePacks}, the core routine responsible for processing
fixed-size data chunks during collective execution.
This routine loads elements from source buffers, applies the required
reduction operator (e.g., sum, min, or max), and writes the results
back to global memory.
We fuse compression and decompression into this stage, allowing data to
be compressed immediately before transmission and decompressed upon
arrival.
By eliminating intermediate memory transfers between computation and
communication, compressed data can be transmitted directly over NVLink
or RDMA, reducing memory traffic and improving pipeline efficiency.

\parhead{Warp-level execution and compatibility.}
Compression and communication are executed at warp granularity,
with each warp responsible for processing one compression or
decompression block.
This design matches NCCL’s warp-cooperative execution model, allowing
independent warps to process data segments and issue communication
operations without global synchronization.
The warp-level structure also preserves architectural portability:
because NCCL collectives are already implemented as warp-cooperative
kernels, our approach requires no architecture-specific tuning across
GPU generations.
In practice, compression executes within the same kernel launch as
reduction and communication, enabling effective overlap between computation and data transfer while utilizing otherwise idle SM cycles.

\parhead{Amortizing compression metadata overhead.}
To amortize metadata overhead, we transmit the ANS table only once per collective invocation.
Float tensors typically exhibit stable value distributions across communication steps, rendering per-block table recomputation unnecessary.
We therefore track the initialization state of compression tables and transmit the table only during the first step of the collective, reusing it thereafter.
This design renders metadata overhead negligible for large-scale collectives.

\parhead{Intra- vs.\ inter-node communication.}
For intra-node communication, compressed data is transmitted directly between GPUs via NVLink.
For inter-node communication over RDMA, as well as intra-node transfers over PCIe, compressed outputs are written to a GPU-resident FIFO buffer, which is consumed by the host networking stack and forwarded through the network interface.
This design eliminates additional data copies between compressed buffers and NCCL buffers, while preserving compatibility with NCCL’s existing transport abstractions.

\parhead{Selective compression across collective stages.}  
Not all data in collective operations benefit equally from compression. 
For point-to-point primitives such as \texttt{send} and \texttt{recv}, we always apply compression before transmission and decompression upon reception. 
However, multi-step collectives such as \texttt{all\_reduce} require finer control because data passes through several intermediate aggregation stages, each potentially involving multiple sources and destinations.  

We therefore apply compression selectively:
on the sender side, compression is performed only when a GPU transmits data to another node or writes to a remote FIFO buffer.  
During intermediate reduction steps that combine locally available data with received partial results, only the remote data are decompressed and merged, while local GPU data remain uncompressed.  
This policy minimizes unnecessary computation while ensuring that all cross-node traffic is transmitted in compressed form.  
Similarly, on the receiver side, decompression is triggered only for data that originates from remote peers, while final output paths bypass decompression entirely.  

\parhead{Handling data alignment.}
NCCL processes tensors in fixed-size chunks.
To simplify integration and avoid partial-block handling overhead,
compression is applied only to chunk-aligned regions (e.g., 32\,KB).
Any remaining tail elements are transmitted in uncompressed form.
This selective strategy preserves correctness across heterogeneous
message sizes while avoiding additional synchronization or control
logic.

\section{Implementation}

We integrate \sysp{} into two GPU communication systems, UCCL-P2P~\cite{uccl_p2p} and NCCL~\cite{nccl}, referred to as \sysp{} and \sysnccl{}, respectively. \sysp{} supports both NVIDIA and AMD GPU platforms, enabling cross-vendor applicability. Specifically, we extend NCCL v2.23.4 with approximately 1.8\,K lines of CUDA code to incorporate compression, and integrate compression into UCCL-P2P with around 4\,K lines of C++ code. Notably, our modifications are lightweight and modular, and can be readily ported to RCCL (the AMD counterpart of NCCL) with minimal effort. 
In both systems, the integration is fully transparent to user applications.

\subsection{Peer-to-Peer Communication Implementation}
We build \sysp on UCCL-P2P, providing synchronous and asynchronous send/recv with transparent compression. UCCL-P2P offers easy-to-use P2P APIs and achieves performance comparable to or exceeding NCCL/RCCL~\cite{nccl,rccl}, NVIDIA NIXL~\cite{nixl2025}, and Mooncake TE~\cite{mooncake_te}.
Our design requires no modifications to the user-facing API. 

Compression changes the effective size of transmitted data, which requires modifications to the metadata management in UCCL-P2P. 
UCCL-P2P is built on top of the RDMA \texttt{write\_with\_imm} primitive provided by RDMA-capable NICs, which enables one-sided remote memory writes without CPU involvement on the receiver side. 
This operation requires the sender to obtain the remote memory address and size in advance.
UCCL-P2P allocates a fixed-size metadata buffer at initialization to coordinate data addresses and sizes for \texttt{write\_with\_imm} during send and receive. To support compression, \sysp{} extends the metadata to include additional information such as the data type and the sizes before and after compression, enabling correct reconstruction and efficient transmission of compressed data.

We extend the DietGPU~\cite{dietgpu_github} compression implementation to support both NVIDIA and AMD GPUs, and leverage UCCL-P2P’s support for diverse RDMA-capable NICs (e.g., EFA and InfiniBand) to enable seamless deployment across heterogeneous GPU–NIC combinations without requiring application changes. We replace PTX-dependent components in DietGPU with portable, HIP-compatible implementations to support AMD GPUs while preserving performance-critical optimizations.

We extend DietGPU~\cite{dietgpu_github} to support multiple floating-point formats. 
\sysp supports the primary data types used in modern LLM, including \texttt{float16}, \texttt{float32}, \texttt{bfloat16}, \texttt{float8\_e4m3fn}, and \texttt{float8\_e5m2}. In particular, we extend support to both FP8 formats. We pack two FP8 values into a single 16-bit unit and jointly extract their exponent fields, producing an 8-bit exponent stream for compression. This design enables byte-granular writes in the split stage and avoids memory misalignment overhead. 

We employ a singleton \textit{Compressor} to orchestrate the compression pipeline. 
The compressor maintains GPU-resident buffers for both compression and decompression. 
It exposes a set of modular primitives corresponding to distinct stages of the compression workflow. Given that UCCL-P2P uses a single send thread and a single receive thread per GPU, we adopt a single compressor instance per GPU to serve all operations, thereby avoiding redundant buffer allocation and reducing memory overhead. 
Combined with chunk-based transmission for large messages, this design bounds the total memory footprint to $\sim$164\,MB per GPU, which accounts for only about 0.1\% of HBM capacity on an H200 GPU.

\subsection{NCCL Compression Implementation}

\sysnccl{} integrates \sys{} into NCCL v2.23.4~\cite{nccl} through lightweight modifications to the collective communication pipeline. 
The implementation adds ~1.8K lines of CUDA to NCCL, embedding compression directly into collective kernels to enable transparent support for Ring \texttt{all\_reduce}, \texttt{all\_to\_all},
and \texttt{send/recv} without API changes.

\sysnccl{} performs compression entirely within the collective kernels using a shared-memory workspace. 
Each kernel allocates a contiguous scratch region to store intermediate state required by the compression pipeline, including histograms, probability tables, symbol buffers, and encoded outputs. 
The workspace is logically partitioned into sender-side compression buffers and receiver-side decompression buffers. 
During transmission, the sender constructs histograms, generates ANS encoding tables, and produces compressed blocks, while the receiver reuses the same region to reconstruct decoding tables and decompress incoming symbols. 
This design enables fully in-kernel compression and decompression, eliminating additional global memory staging and auxiliary kernel launches.

\sysnccl{} integrates compression into the collective data path via a warp-parallel pipeline that operates on fixed-size blocks in each iteration of the collective loop. 
Each value is decomposed into a non-compressible component (sign and exponent), which is transmitted directly, and a compressible mantissa symbol used for encoding. 
Threads cooperatively build histograms over mantissa symbols to construct ANS probability tables, after which each warp independently encodes its local symbols into a variable-length compressed stream. 
The resulting compressed data and minimal metadata are written directly into the communication buffers, replacing the original data blocks in the collective pipeline.

\section{Evaluation}
\label{sec:eval}

\begin{table}[t]
\centering
{\small
\setlength{\tabcolsep}{2pt}
\begin{tabular}{lccc}
\toprule
Cluster & GPU & CPU (\#cores) & Network \\
\midrule
AWS EFA & 8$\times$H200 & Xeon 8488C (192) & 16$\times$EFA (200\,Gbps) \\
AMD & 8$\times$MI355X & EPYC 9575F (128) & 8$\times$RDMA NICs (400\,Gbps) \\
AWS g6e & 8$\times$L40S & 192 vCPUs &  Ethernet (400\,Gbps) \\
\bottomrule
\end{tabular}
}
\caption{Hardware specifications of our experimental platforms. }
\label{tab:testbeds}
\end{table}

\subsection{Experiment Setup}
\textbf{Testbeds.}
We evaluate \sys across diverse GPU–NIC platforms, including two \texttt{p5en.48xlarge} instances connected via Elastic Fabric Adapter (EFA), a two-node AMD cluster with AMD Instinct MI355X GPUs connected via RoCEv2, and a single-node AWS \texttt{g6e.48xlarge} instance with 8 NVIDIA L40S GPUs. By default, \sysp is evaluated on the \texttt{p5en.48xlarge} cluster, while \sysnccl is evaluated on the \texttt{g6e.48xlarge} instance. Hardware specifications are summarized in Table~\ref{tab:testbeds}. \lam{i think it need some justification for two different testbed.. given that the y axis of the (a) and (b) diff is too high, reader will catch it and got confused immediately}

\noindent\textbf{Baselines.}
We compare against multiple baselines. For NCCL with compression, we use NCCL v2.23.4 with its default configuration as the baseline. For fair comparison, NCCL is restricted to 4 SMs, matching \sysnccl. 
For P2P communication, we use UCCL-P2P~\cite{uccl_p2p} with its default synchronous \texttt{send}/\texttt{recv} interface as the baseline, as \sysp is built directly on top of it.
We compare \sysp against two designs: (1) a naive \textit{encode-send} scheme that transmits only after full compression, and (2) a conventional \textit{native pipeline} that partitions data into 8\,MB chunks and overlaps compression with communication. We adopt an 8\,MB chunk size, as it empirically provides a reasonable balance between compression effectiveness and pipeline overhead.

Compression is enabled only for messages larger than 1\,MB, as overhead outweighs benefits for smaller messages. We use bfloat16 as the default data format.

\subsection{End-to-End throughput}
\subsubsection{\sysp Performance.}
\begin{figure}[t]
\centering
\begin{subfigure}[t]{0.45\columnwidth}
    \centering
    \includegraphics[width=\linewidth]{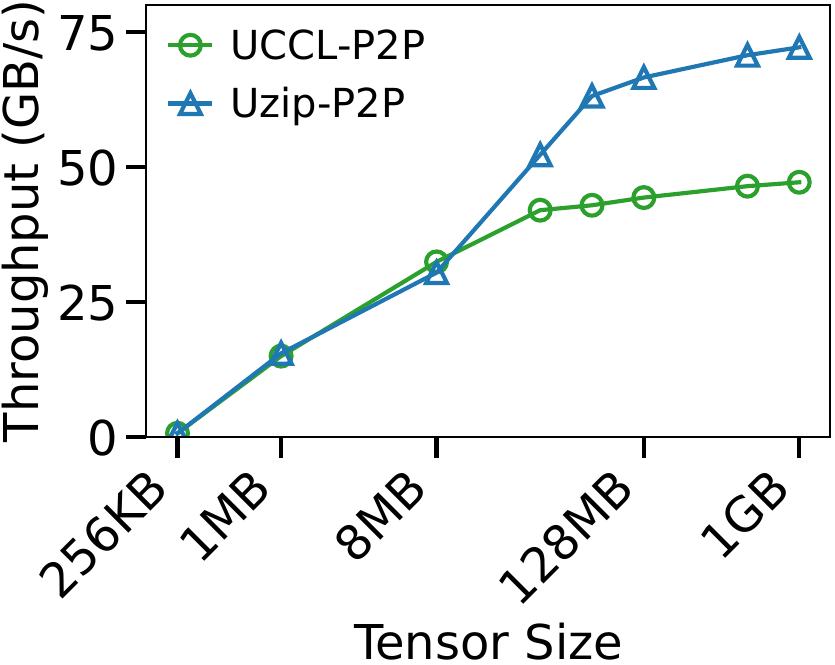}
    \caption{\sysp P2P throughput.}
    \label{fig:throughput_p2p}
\end{subfigure}
\hfill
\begin{subfigure}[t]{0.45\columnwidth}
    \centering
    \includegraphics[width=\linewidth]{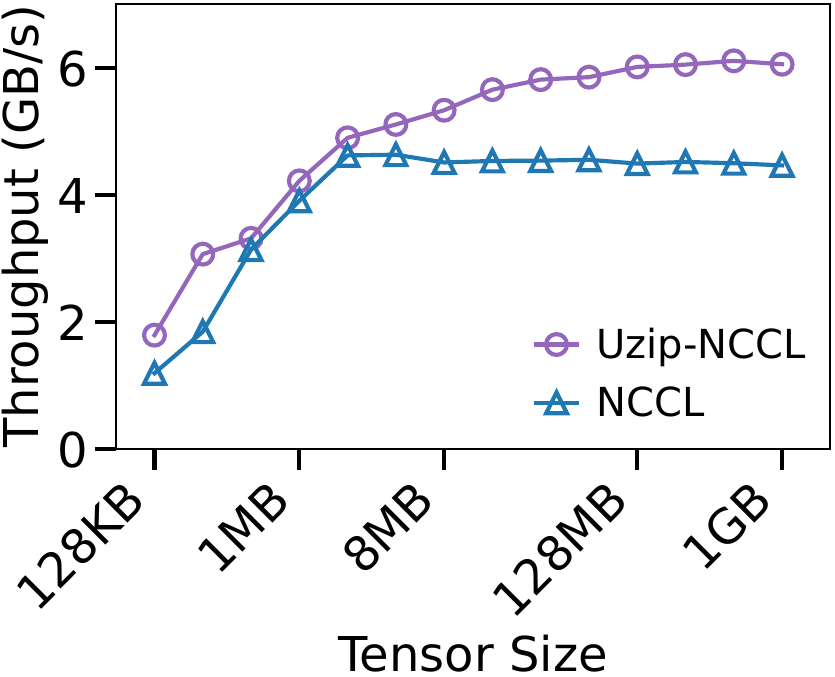}
    \caption{\sysnccl P2P throughput.}
    \label{fig:throughput_nccl_p2p}
\end{subfigure}
\caption{Throughput comparison for bfloat16 peer-to-peer communication across tensor sizes. }
\label{fig:p2p_throughput}
\end{figure}

Figure~\ref{fig:throughput_p2p} reports the throughput of our \sysp compression pipeline across tensor sizes from 256\,KB to 1\,GB. As tensor size increases, \sysp delivers progressively higher performance gains over UCCL-P2P.

For medium-to-large tensors ($>$8\,MB), \sysp consistently outperforms the UCCL-P2P. At 1\,GB, it improves throughput by up to 52.9\% (72.2\,GB/s vs.\ 47.2\,GB/s), closely approaching the theoretical upper bound of 73.8\,GB/s derived from Amdahl’s Law under a 64\% compression ratio. For smaller tensors, the benefits are more modest (e.g., 8\% at 16\,MB and 24\% at 32\,MB), as compression overhead partially offsets bandwidth savings.

We generate \texttt{bf16} tensors with values uniformly distributed in $[-1,1]$, achieving a stable compression ratio of approximately 64\% across all tensor sizes. \done{(Lam)we dont have real data distribution? if no, we need to justify this range is reasonable, or if this 64\% is a upper bound, you need to say it upfront in this subsection Shuang: we have later, close to 0.64} For small messages (e.g., 4\,MB), compression overhead cannot be fully amortized. However, in practical workloads such as RL weight updates and KV cache transfers, tensor sizes typically exceed this threshold, making \sysp effective in practice.

\begin{figure}[t]
\centering
\begin{subfigure}[t]{0.45\columnwidth}
    \centering
    \includegraphics[width=\linewidth]{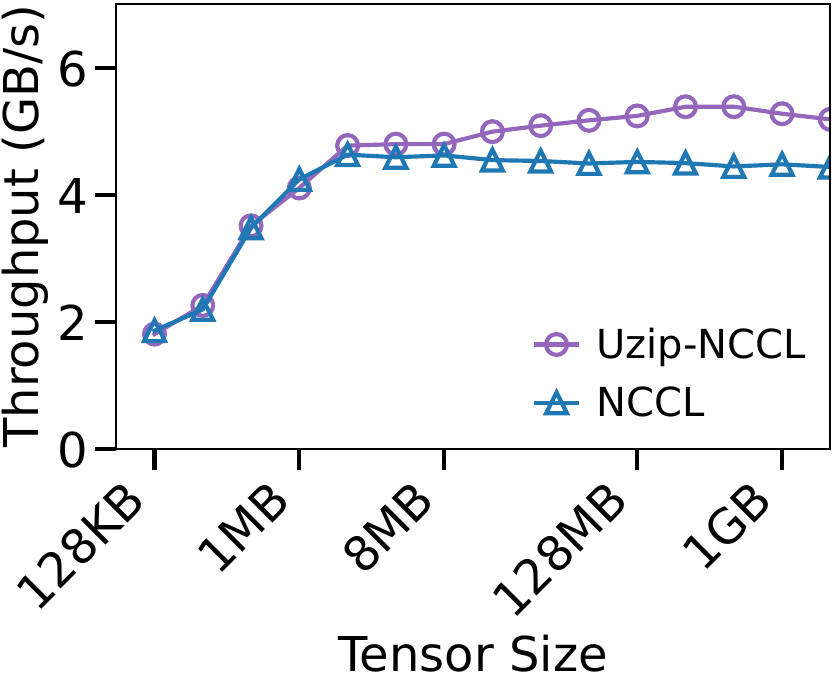}
    \caption{\texttt{all\_to\_all} throughput.}
    \label{fig:all_to_all_throughput}
\end{subfigure}
\hfill
\begin{subfigure}[t]{0.45\columnwidth}
    \centering
    \includegraphics[width=\linewidth]{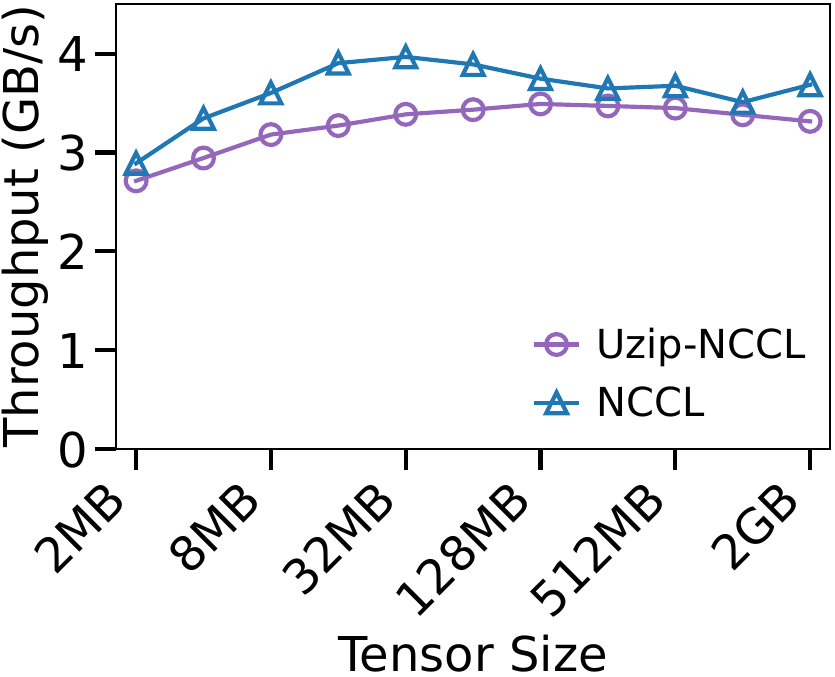}
    \caption{Ring \texttt{all\_reduce} throughput.}
    \label{fig:all_reduce_throughput}
\end{subfigure}
\caption{Throughput of \sysnccl collective communication primitives  across varying sizes.}
\label{fig:collective_throughput}
\end{figure}

\subsubsection{\sysnccl Performance.}
We evaluate the performance of \sysnccl on two representative collective
operations widely used in NCCL: \texttt{all\_to\_all} and
\texttt{all\_reduce}. We also evaluate the point-to-point
\texttt{send/recv} primitive. 

The \sysnccl consistently
outperforms NCCL for \texttt{send/recv} across all message
sizes (Figure~\ref{fig:throughput_nccl_p2p}), achieving over 20\% higher throughput for messages larger than
32\,MB. A similar trend is observed for the \texttt{all\_to\_all} collective
(Figure~\ref{fig:all_to_all_throughput}), which effectively consists
of multiple \texttt{send/recv} operations. Due to additional GPU scheduling and synchronization overheads, the performance gain is slightly reduced, reaching about 18\% for large messages ($>$32\,MB).

\begin{figure}[t]
\centering
\begin{subfigure}[t]{0.48\columnwidth}
    \centering
    \includegraphics[width=\linewidth]{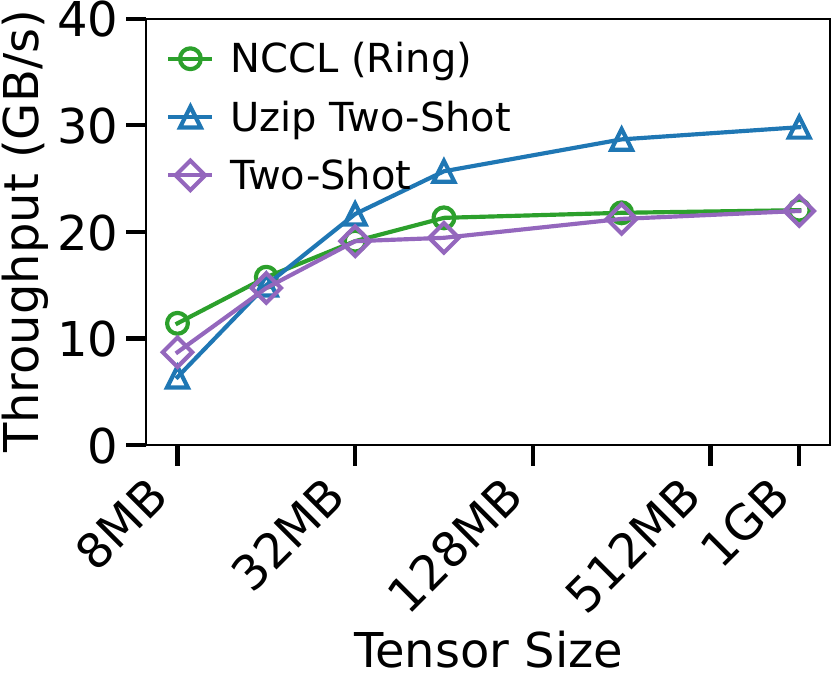}
    \caption{Two-Shot \texttt{all\_reduce} throughput.}
    \label{fig:throughput_two_shot}
\end{subfigure}
\hfill
\begin{subfigure}[t]{0.48\columnwidth}
    \centering
    \includegraphics[width=\linewidth]{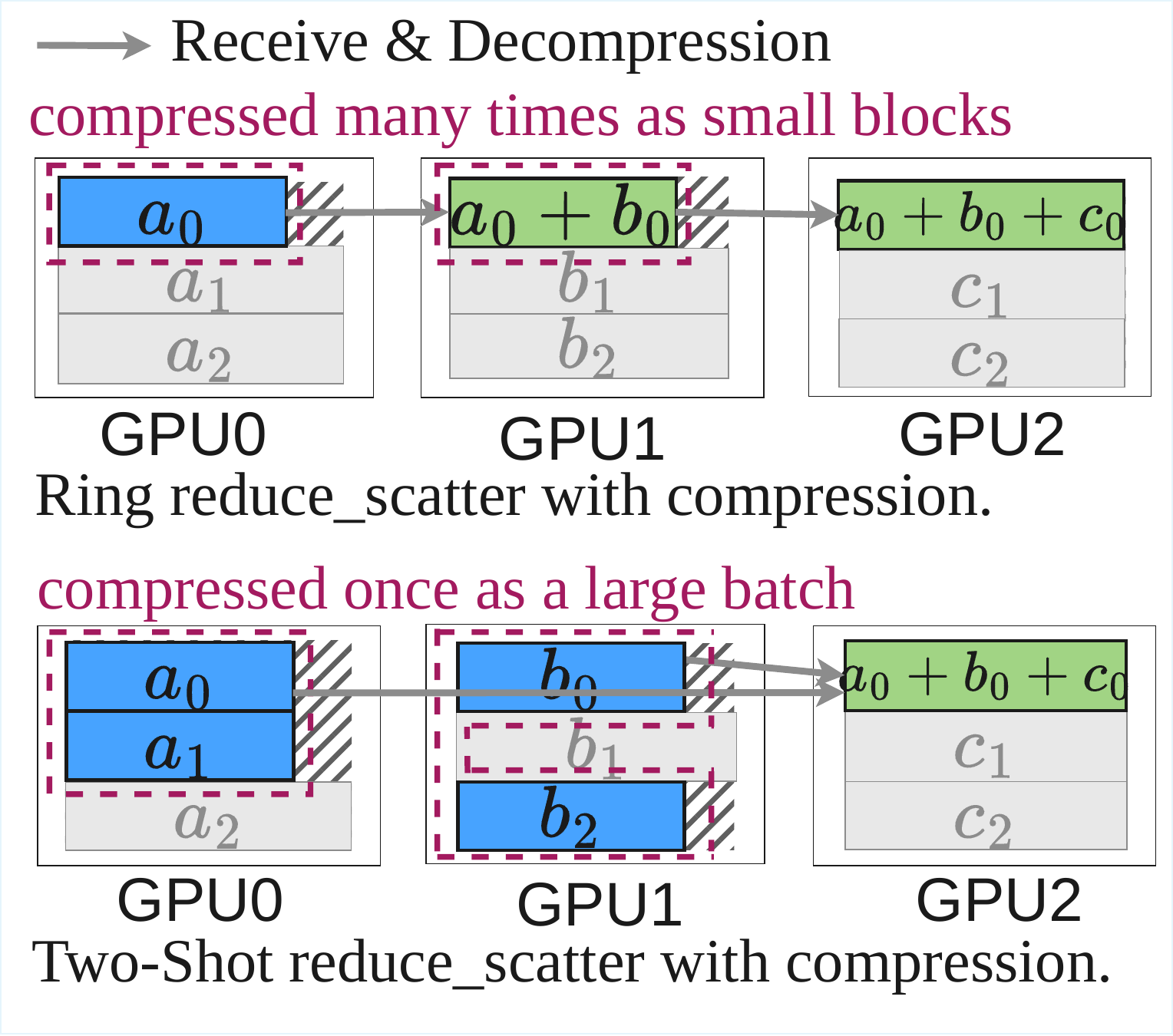}
    \caption{Two-Shot vs. Ring \texttt{all\_reduce}\yang{I think two-shot will send a1 and b1 to c1. Better to draw how three sets of chunks send data among them. Better to put ring AR first. You also need to somehow highlight that the numbers of compressions and decompressions are different. Then mention the figure shows the reduce\_scatter phase, and later all\_gather phase is similar.}.}
    \label{fig:compare_two_shot}
\end{subfigure}
\caption{Throughput of two-shot \texttt{all\_reduce} implemented with asynchronous \texttt{isend}/\texttt{irecv} on two \texttt{p5en.48xlarge} instances (16 GPUs total) with NVLink disabled.}
\label{fig:collective_throughput_two_shot}
\end{figure}

\sysnccl performs worse than NCCL for the
\texttt{all\_reduce} (Figure~\ref{fig:all_reduce_throughput}).
This is because NCCL adopts a ring-based \texttt{all\_reduce} algorithm, which is inherently unfriendly to compression integration. \texttt{all\_reduce} consists of two phases: reduce-scatter and all-gather. The all-gather phase involves only data movement without computation, and thus exhibits similar compression behavior across different implementations (e.g., ring-based and two-shot).  In contrast, the reduce-scatter interleaves communication with reduction operations (e.g., sum or average), which imposes stricter constraints on compression. Each received data chunk must be decompressed before the reduction. \lam{i am confused, 8a is not ring? whats that} As shown in Figure~\ref{fig:compare_two_shot}, the ring-based implementation partitions data into small chunks and propagates them along a logical ring topology. Each chunk undergoes multiple rounds of compression, transmission, decompression, and reduction as it traverses the ring. 
This repeated compression–decompression cycle incurs significant overhead, reducing the effectiveness of compression in NCCL \texttt{all\_reduce}.

Instead, the two-shot \texttt{all\_reduce} design significantly reduces this overhead.\ziming{What is the two-shot here? I might have missed it} \lam{i think we should change a way to bring out this two-shot all reduce, not this two-shot reduce the overhead, its about two-shot fits our compression where ring does not fit, make it a design choice not a compromise} Each GPU compresses its data only once as a large chunk or batch before transmission, and the receiving GPU performs a single decompression prior to reduction (Figure~\ref{fig:compare_two_shot}).
We implement a two-shot \texttt{all\_reduce} using asynchronous \texttt{isend}/\texttt{irecv} primitives and integrate \sys{} into the communication path (Figure~\ref{fig:throughput_two_shot}).
Unlike NCCL's ring-based \texttt{all\_reduce}, which partitions data into fine-grained per-CTA slices, our implementation operates on large contiguous chunks. This coarse-grained design, combined with the reduced number of compression invocations inherent to the two-shot approach, allows compression to be applied more efficiently. \done{You need to explain that NCCL uses ring-based AR, and then explain the difference between two-shot and ring-based \texttt{all\_reduce}, so people can understand why two-shot gives better performance.}\done{(Yang) I guess here we want to mention that our self-implemented allreduce does not do fine-grained chunking, thus having large chunks for compression/decompression, and two-shot also gives larger chunk size for compression. }\sys Two-Shot begins to deliver performance benefits once the data size reaches 32\,MB, achieving a 13.3\% throughput improvement over the baseline two-shot implementation. The gain continues to increase with larger message sizes, reaching up to 35.7\% at 1\,GB.

\begin{figure}[htbp]
\centering
\begin{subfigure}{0.90\linewidth}
\centering
\includegraphics[width=\linewidth]{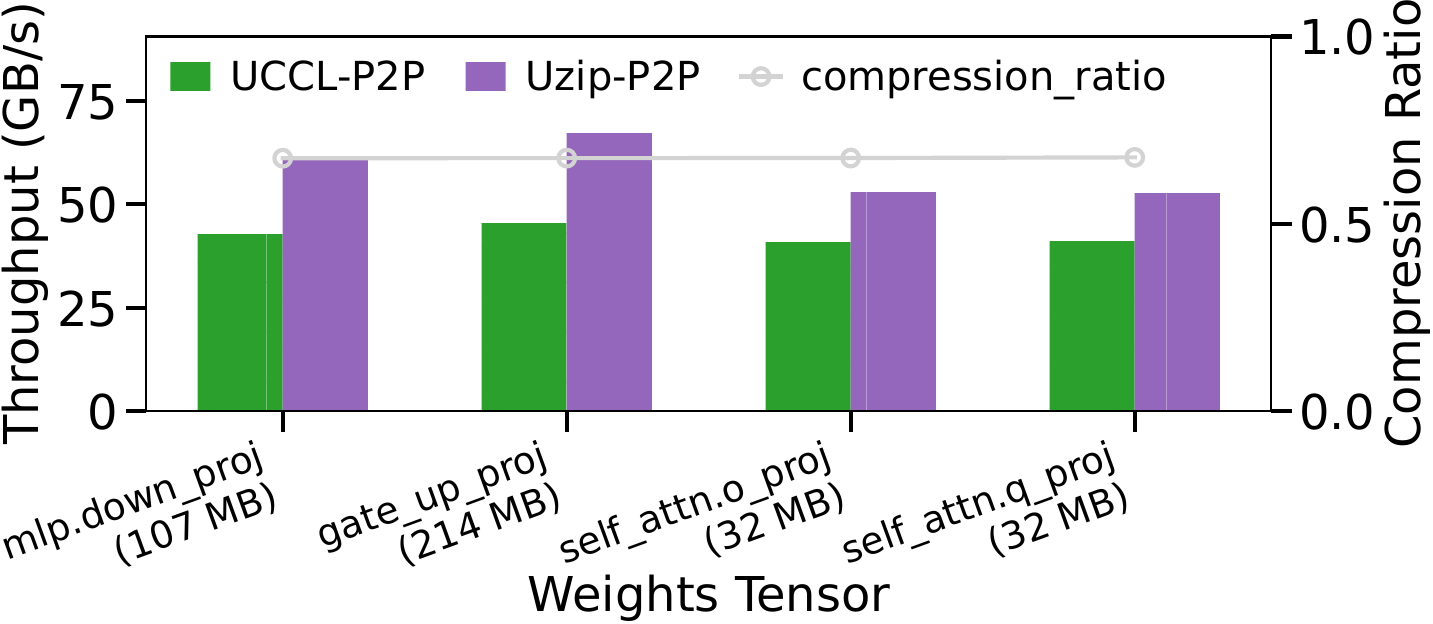}
\caption{Weight update throughput for GLM4-9B.}
\label{fig:weight_update_exp_GLM}
\end{subfigure}
\begin{subfigure}{0.90\linewidth}
\centering
\includegraphics[width=\linewidth]{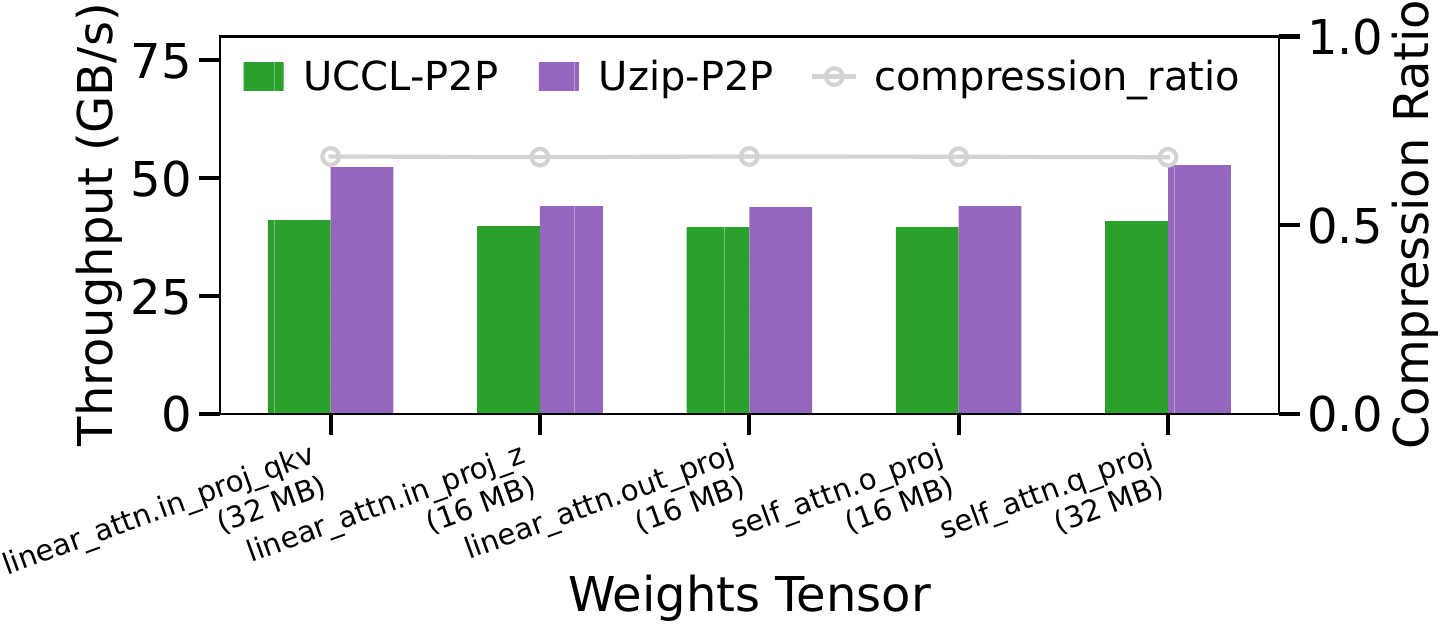}
\caption{Weight update throughput for Qwen3.5-35B-A3B.}
\label{fig:weight_update_exp_Qwen}
\end{subfigure}
\caption{Application-level evaluation of \sysp on bf16 weight tensors during RL training. (a) GLM4-9B and (b) Qwen3.5-35B-A3B. }
\label{fig:weight_update_exp}
\end{figure}

\begin{figure}[t]
\centering
\includegraphics[width=0.90\columnwidth]{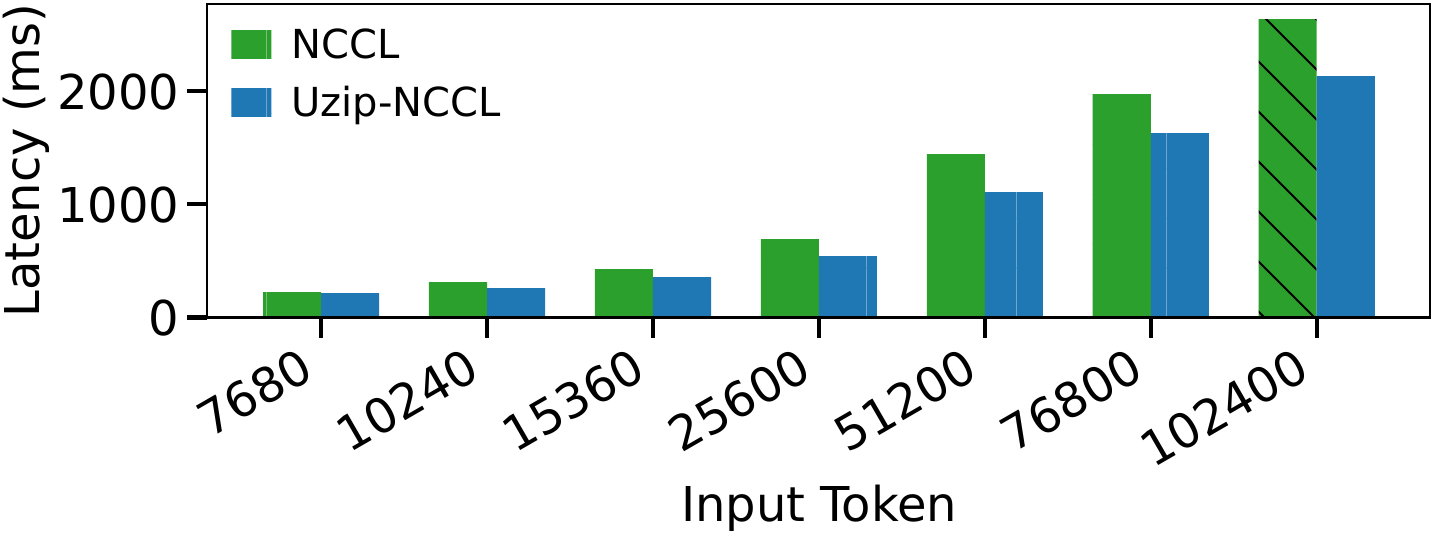}
\caption{KV cache transfer latency under prefill–decode disaggregation (P1D3) in vLLM for the Qwen-7B-Chat model.}
\label{fig:pd_Disaggregation_exp}
\end{figure}

\vspace{-10pt}
\subsection{Application-Level Performance}
\subsubsection{Weight Updates in Reinforcement Learning.}
We evaluate \sysp on weight tensors generated during the update phase of LLM reinforcement learning. We consider two representative models: the dense GLM4-9B (9B parameters) and the mixture-of-experts (MoE) model Qwen3.5-35B-A3B (35B parameters). The training pipeline runs on 8 GPUs, where 4 GPUs perform policy optimization and the remaining 4 GPUs generate rollouts.

Figure~\ref{fig:weight_update_exp} presents the communication throughput across representative weight tensors from different layers and training stages. The x-axis lists tensor names and sizes. 

For the dense GLM4-9B model (Figure~\ref{fig:weight_update_exp_GLM}), weight tensors are relatively large, enabling substantial communication reduction. In particular, \sysp achieves up to 47.5\% higher throughput for the \texttt{gate\_\allowbreak up\_\allowbreak proj} tensor (214\,MB).

For the MoE model Qwen3.5-35B-A3B (Figure~\ref{fig:weight_update_exp_Qwen}), tensors are smaller due to sparse expert activation. Despite this, \sysp still delivers up to 28.8\% improvement for the \texttt{self\_attn.q\_proj.weight} tensor (32\,MB). For smaller tensors (e.g., 16\,MB), \sysp maintains around 10\% improvement. Finally, the compression ratio remains stable across models, layers, and training stages. As it directly determines communication reduction, this stability indicates that our lossless design generalizes well to diverse LLM training workloads.

\subsubsection{KV Cache Transfer in Prefill–Decode Disaggregation.}
We integrate \sysnccl with the Prefill–Decode disaggregation inference pipeline of vLLM~\cite{kwon2023efficient} to evaluate its performance in realistic distributed LLM serving workloads \emph{without application changes}.
Experiments follow the default Prefill–Decode disaggregation
configuration (P1D3) in vLLM, where one GPU performs the prefill
stage and three GPUs execute decoding. We measure the KV cache transfer latency. Figure~\ref{fig:pd_Disaggregation_exp} shows that \sysnccl consistently
reduces KV cache transfer latency compared to NCCL, achieving up to
30.1\% improvement.
Since KV cache transmission constitutes a significant portion of
the inference latency, these improvements translate directly into
end-to-end performance gains. For example, when the input length is
7,680 tokens, KV cache transfer accounts for approximately
23\% of the total execution time, implying an overall application
speedup of about 10\% over the NCCL.

\subsection{Ablation Studies}
\subsubsection{Transmission Throughput and Compression Ratio Across RL Steps}
\begin{figure}[t]
\centering
\includegraphics[width=0.9\columnwidth]{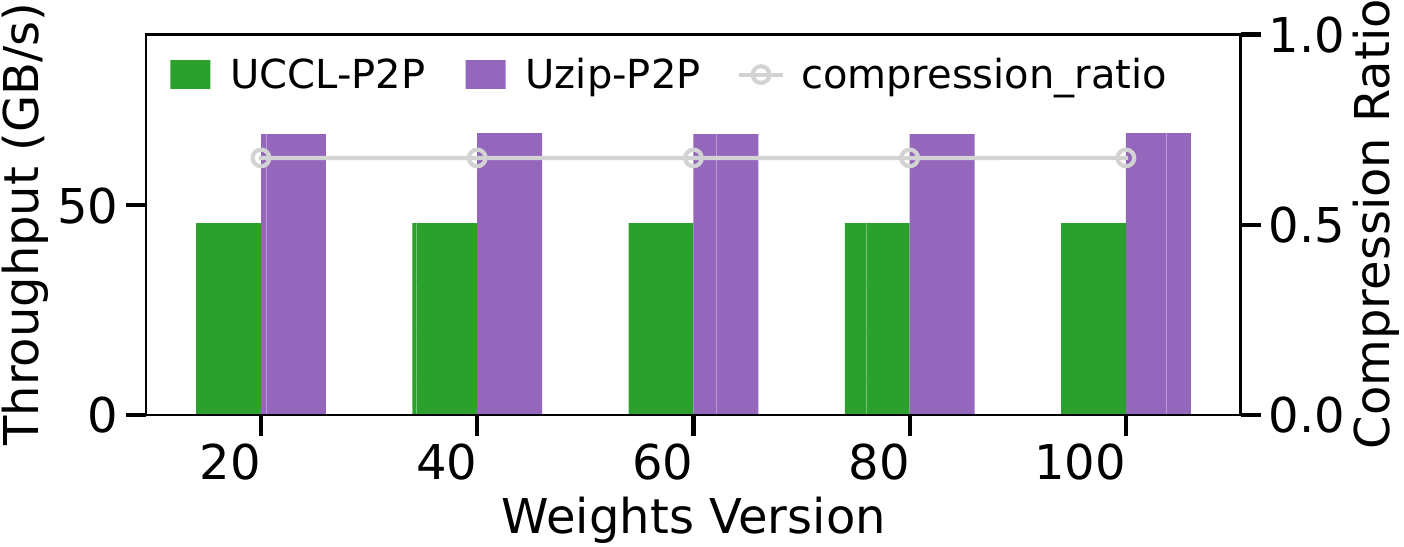}
\caption{Communication throughput for different versions of the
\texttt{gate\_up\_proj} weight tensor (214\,MB) during GLM4-9B
RL training.}
\label{fig:weights_update_diff_version_exp}
\end{figure}

We examine compression behavior during reinforcement learning training by measuring throughput and compression ratio for different versions of the \texttt{gate\_up\_proj} weight tensor (214\,MB) from GLM4-9B. As shown in Figure~\ref{fig:weights_update_diff_version_exp}, the compression ratio remains stable across training checkpoints, close to that of randomly generated tensors. Correspondingly, \sysp consistently outperforms the UCCL-P2P, achieving stable throughput gains throughout training.

\begin{figure}[t!]
\centering
\begin{subfigure}[t]{0.48\columnwidth}
    \centering
    \includegraphics[width=\linewidth]{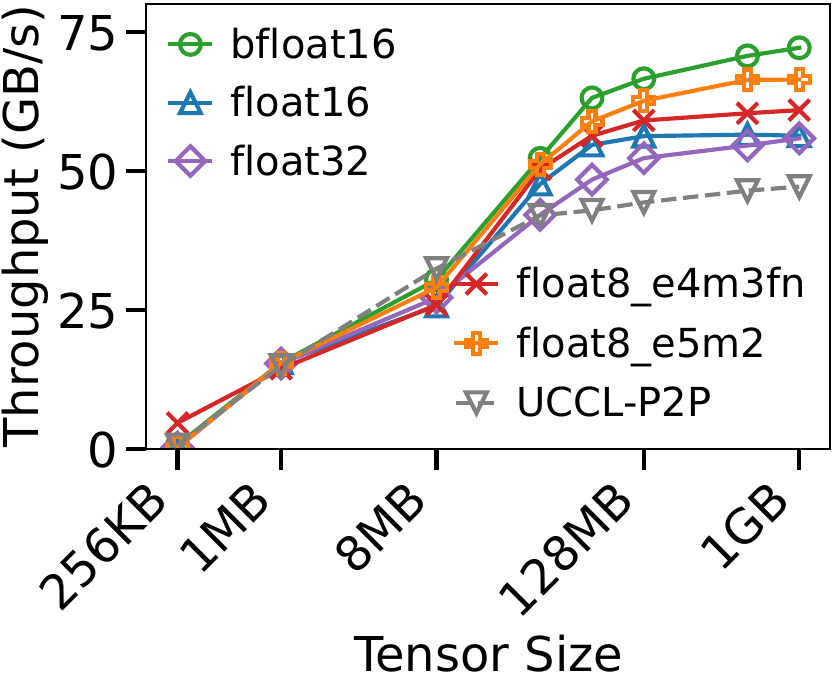}
    \caption{\sysp throughput across floating-point types.}
    \label{fig:different_float_types_a}
\end{subfigure}
\hfill
\begin{subfigure}[t]{0.48\columnwidth}
    \centering
    \includegraphics[width=\linewidth]{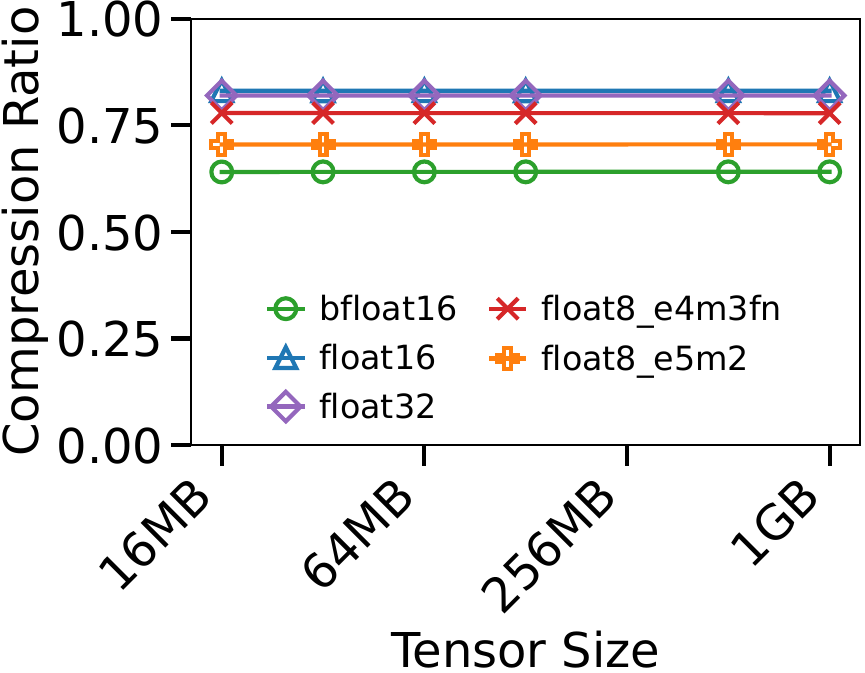}
    \caption{Compression ratios across floating-point types.}
    \label{fig:different_float_types_b}
\end{subfigure}
\caption{Performance of \sysp across different floating-point
data types.}
\label{fig:different_float_types}
\end{figure}

\subsubsection{Performance Across Different Floating-Point Data Types.}
We evaluate \sysp across different floating-point formats, including \texttt{float16}, \texttt{float32}, \texttt{float8\_e4m3fn}, and \texttt{float8\_e5m2} (Figure~\ref{fig:different_float_types}). 
The achievable compression ratio is primarily determined by the floating-point exponent field, as our scheme encodes only exponent values. In IEEE formats, \texttt{float16} uses 5 exponent bits out of 16 bits, while \texttt{float32} uses 8 exponent bits out of 32 bits. Consequently, both formats offer limited compression opportunities compared to \texttt{bfloat16}, which retains an 8-bit exponent within a 16-bit representation. This trend is reflected in Figure~\ref{fig:different_float_types_b}, where the compression ratios for \texttt{float16}, \texttt{float32}, \texttt{bfloat16}, \texttt{float8\_e4m3fn}, and \texttt{float8\_e5m2} are approximately 83\%, 82\%, 64\%, 77\%, and 70\%, respectively.

These differences translate into performance gains in Figure~\ref{fig:different_float_types_a}. \sysp achieves up to 41.9\% throughput improvement for \texttt{float8\allowbreak\_e5m2}, compared to 30.2\% for \texttt{float8\_e4m3fn} over UCCL-P2P.

\vspace{-8pt}
\begin{figure}[t]
\centering
\begin{minipage}[t]{0.48\columnwidth}
    \centering
    \includegraphics[width=\linewidth]{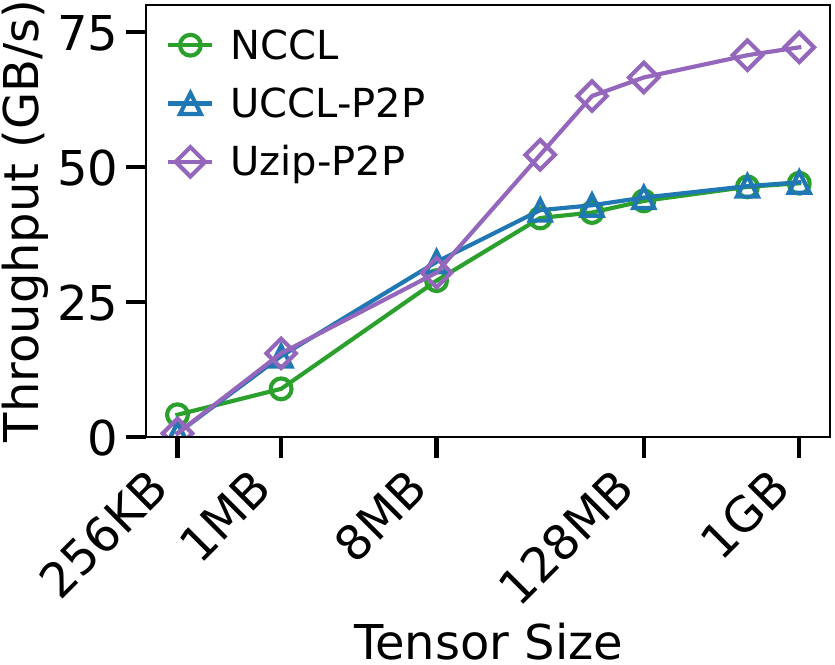}
    \captionof{figure}{\sysp throughput comparison with NCCL \texttt{send/recv}.}
    \label{fig:nccl_baseline_split_send}
\end{minipage}
\hfill
\begin{minipage}[t]{0.48\columnwidth}
    \centering
    \includegraphics[width=\linewidth]{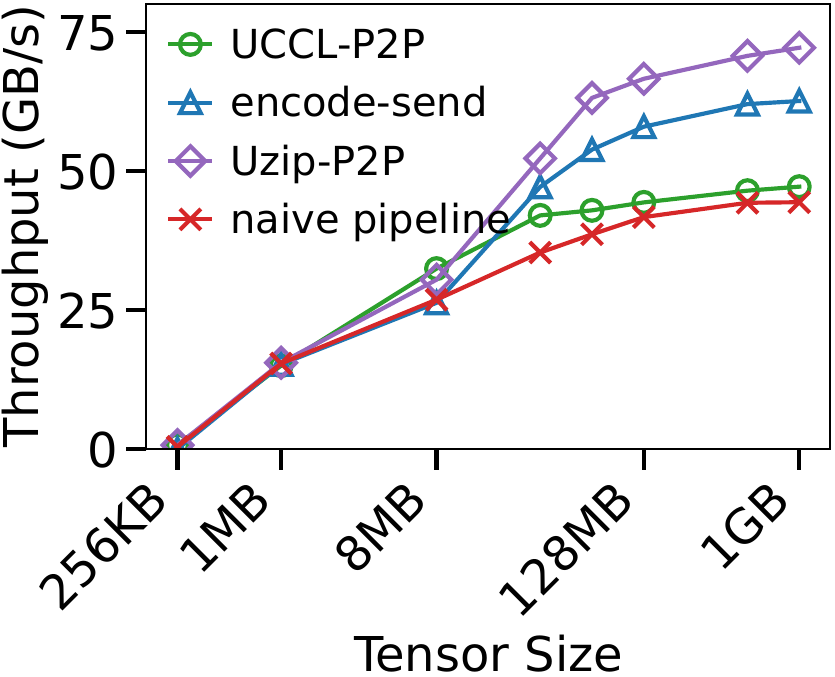}
    \captionof{figure}{Throughput of compression--communication integration strategies.}
    \label{fig:compare_with_native_pipeline}
\end{minipage}
\end{figure}

\begin{figure}[!t]
\centering
\begin{subfigure}[t]{0.48\columnwidth}
    \centering
    \includegraphics[width=\linewidth]{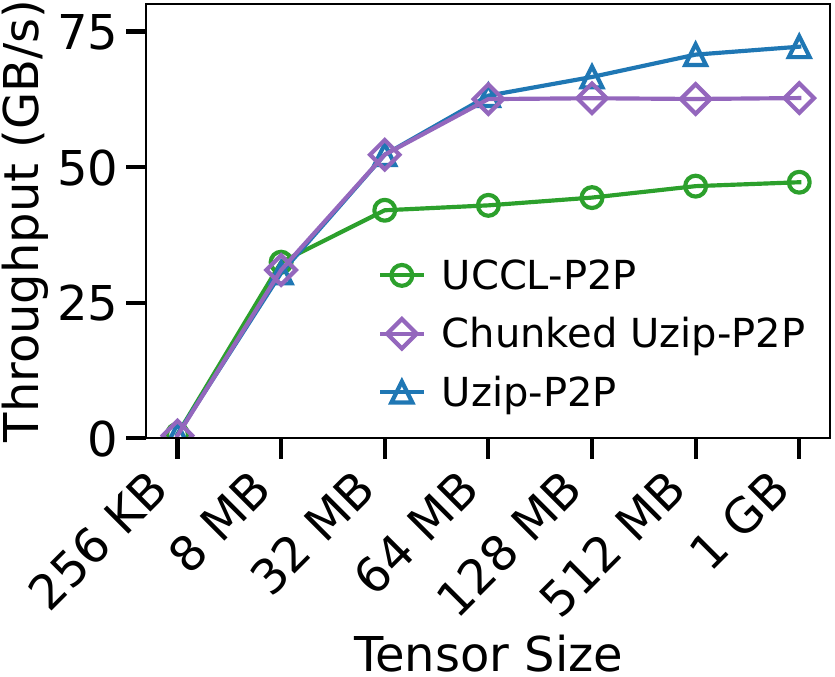}
    \caption{Impact of buffer memory footprint.}
    \label{fig:resource_usage_memory}
\end{subfigure}
\hfill
\begin{subfigure}[t]{0.48\columnwidth}
    \centering
    \includegraphics[width=\linewidth]{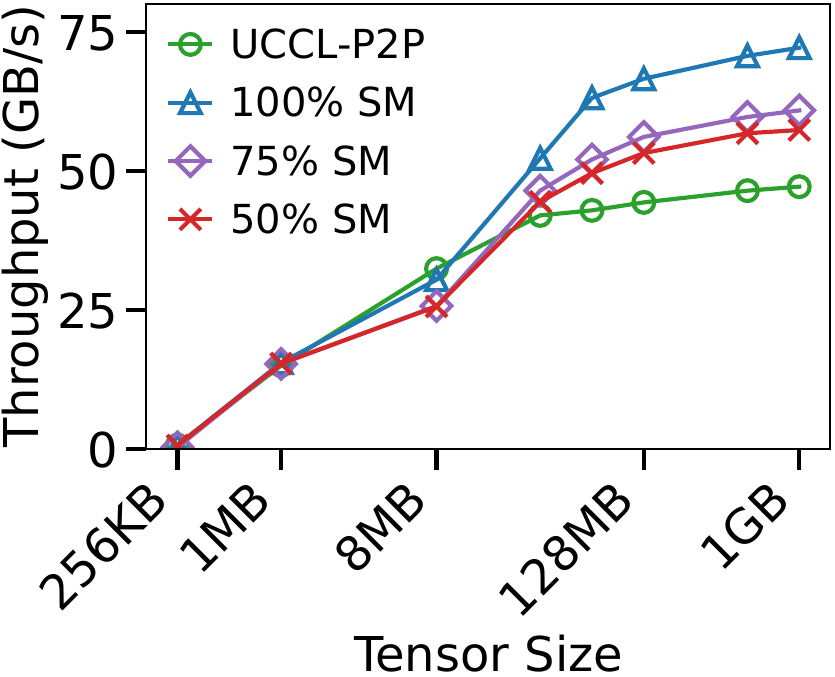}
    \caption{Impact of SM utilization constraints.}
    \label{fig:resource_usage_sm}
\end{subfigure}
\caption{Throughput of \sysp under constrained GPU resources.}
\label{fig:resource_usage}
\end{figure}

\subsubsection{Comparison with NCCL \texttt{send/recv}.}
We further compare baseline UCCL-P2P implementation and \sysp against the NCCL \texttt{send/recv} interface.
 The UCCL-P2P implementation achieves performance comparable to NCCL for tensors larger than 8\,MB and slightly outperforms NCCL for smaller messages (Figure~\ref{fig:nccl_baseline_split_send}).
By incorporating compression, \sysp further improves throughput and consistently outperforms NCCL across the entire range of tensor sizes.

\vspace{-8pt}
\subsubsection{Comparison with Naive Methods.}
We compare \sysp against two alternative designs for integrating compression with communication. 
Figure~\ref{fig:compare_with_native_pipeline} shows that \sysp consistently outperforms both alternatives across tensor sizes. 
For small tensors (e.g., 8\,MB), compression overhead dominates, and \textit{encode-send} degrades throughput by 18\% relative to the UCCL-P2P, while \sysp reduces this to 6\% via partial overlap. 
For larger tensors, \sysp outperforms \textit{encode-send} by effectively overlapping compression and communication. 
The \textit{naive pipeline} slightly underperforms the UCCL-P2P, as fine-grained chunking incurs excess compression overhead that outweighs pipelining benefits.

\vspace{-8pt}
\subsubsection{Performance under Resource Constraints}
We evaluate the robustness of \sysp under constrained GPU memory and limited SM availability. 

\parhead{Memory Footprint.}
\sysp incurs additional GPU memory overhead for staging compression buffers, scaling with block granularity. To reduce memory usage, large tensors can be partitioned into smaller chunks at the cost of extra kernel launches. Despite this trade-off, \sysp remains effective under tight memory constraints: with a 164\,MB buffer per GPU (approximately 0.1\% of the 141\,GB HBM on H200), chunked \sysp{} still achieves up to 41.2\% throughput improvement for 128\,MB tensors (Figure~\ref{fig:resource_usage_memory}).

\parhead{SM Utilization.}
We evaluate \sysp under restricted SM availability using CUDA Green Context~\cite{cui2025optimizingsloorientedllmserving}. As shown in Figure~\ref{fig:resource_usage_sm}, \sysp achieves up to 20.4\% improvement even with 50\% of SMs. \sysp leverages otherwise idle SMs during communication, as dependencies often prevent LLM communication from fully overlapping with computation.

\subsubsection{\sysp Throughput on AMD GPUs and NICs}
\begin{figure}[!t]
\centering
\begin{minipage}[t]{0.48\columnwidth}
    \centering
    \includegraphics[width=\linewidth]{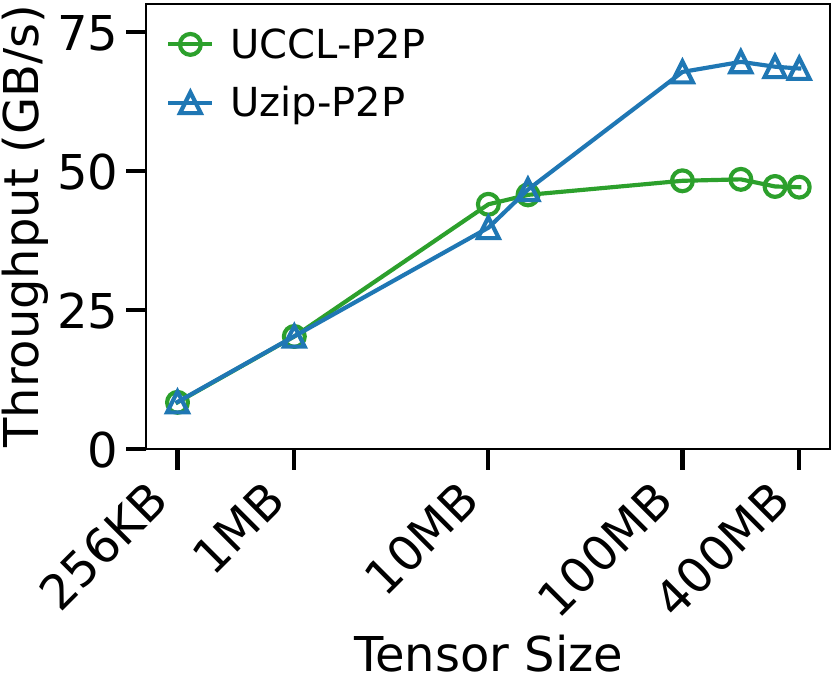}
    \caption{\sysp throughput on AMD MI355X over RoCEv2.}
    \label{fig:amd_p2p_split_vs_2baseline}
\end{minipage}
\hfill
\begin{minipage}[t]{0.48\columnwidth}
    \centering
    \includegraphics[width=\linewidth]{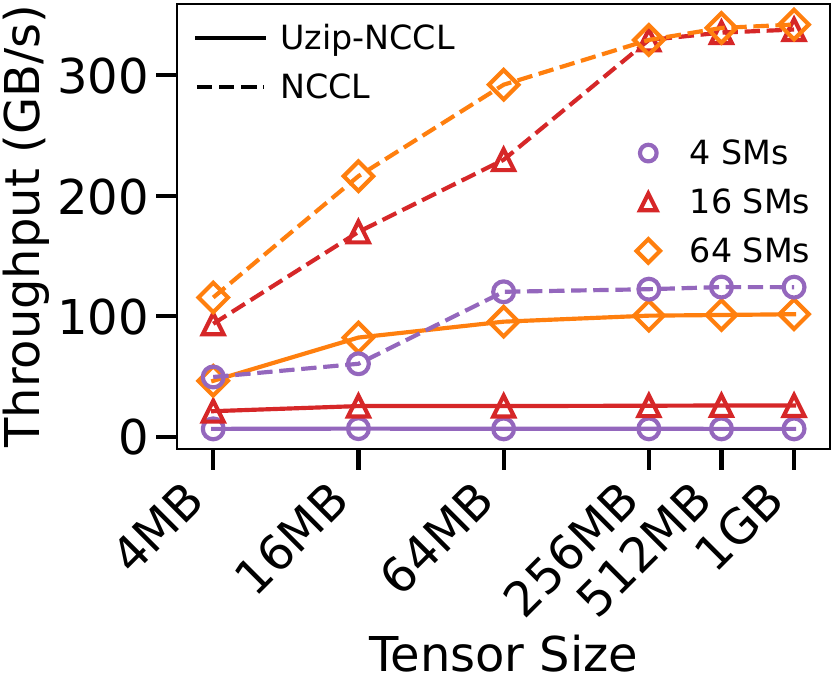}
    \caption{\sysnccl P2P throughput over NVLink on a single p5en.48xlarge node.}
    \label{fig:nvlink_send_recv}
\end{minipage}
\end{figure}

We evaluate \sysp on an AMD cluster with MI355X GPUs and Pensando DSC SmartNICs. Figure~\ref{fig:amd_p2p_split_vs_2baseline} shows that \sysp consistently improves end-to-end P2P throughput across tensor sizes, achieving up to 45.3\% higher throughput at large sizes and generalizing across GPU and NIC vendors.

\section{Discussion}

\parhead{Applying Compression to NVLink Communication.}
We acknowledge that \sysnccl over NVLink gives negative gains, shown in Figure~\ref{fig:nvlink_send_recv}. This is essentially caused by the architecture incompatibility of NCCL with lossless compression: fine-grained chunking causes compression/decompression inefficiency. Addressing this issue would require faster compression algorithm designs and larger compression granularity to better utilize GPU hardware. 

A promising direction is to develop GPU-friendly lossless compression algorithms with more efficient kernel implementation, potentially leveraging Tensor Cores or emerging hardware support for compression primitives~\cite{xu2025llm265}.

\parhead{Building a Compression-Friendly Collective Library.}
Results in Figure~\ref{fig:collective_throughput_two_shot} indicate that reducing the number of compression invocations is critical to amortizing compression overhead. Our two-shot \texttt{all\_reduce} demonstrates the feasibility of compression-aware collective design by minimizing redundant compression and decompression. However, our current implementation is built on NCCL \texttt{isend}/\texttt{irecv}, where reduction and compression remain loosely coupled, incurring additional kernel launches and memory copy overheads.
As future work, we plan to build a compression-friendly collective communication library that tightly integrates compression into collective primitives.

\section{Related Work}
\noindent\textbf{GPU Communication Library.}
Prior work has improved GPU communication efficiency for both P2P and collective operations. Systems such as NIXL~\cite{nixl2025}, Mooncake~\cite{mooncake_te}, pplx-garden~\cite{pplx2025}, and UCCL-P2P~\cite{uccl_p2p} focus on high-performance P2P, while NVSHMEM~\cite{langer2021nvshmem}, NCCL~\cite{nccl}, and RCCL~\cite{rccl} provide both collective primitives and P2P abstractions. More recent efforts, including UCCL~\cite{uccl} and MSCCL++~\cite{msccl}, optimize collective communication at the network and scheduling layers.
However, none of these systems natively support compression within the communication pipeline.
\parhead{Lossless Compression Algorithms on GPUs.}
Existing work explores a variety of lossless GPU compression techniques. ZipNN~\cite{ZipNN2025} employs Huffman-based entropy coding, while NeuZip~\cite{NeuZip2024}, DFloat11~\cite{dfloat11}, TDT~\cite{TDT2025}, and DietGPU~\cite{dietgpu_github} adopt ANS-based designs optimized for GPU execution. MPC~\cite{MPC2015} targets massively parallel architectures, and ZipServ~\cite{fan2026zipserv} proposes hardware-aware compression with fixed-length encoding. Earlier efforts such as Deep Compression~\cite{Han2015DeepCC} apply Huffman coding to quantized models, while nvCOMP~\cite{nvcomp2023} provides general-purpose GPU compression. However, existing approaches largely optimize compression in isolation, without integrating it into GPU communication.


\parhead{GPU Communication Library with Compression.}
Existing work reduces GPU communication volume via quantization and other lossy compression techniques. Systems such as C-Coll~\cite{huang2024mpi}, ZCCL~\cite{zccl2025}, gZCCL~\cite{gzcccl2024}, OmniReduce~\cite{fei2021sparse}, and ghZCCL~\cite{ghzccl2025} apply error-bounded compression to accelerate collectives. Zhou et al.~\cite{zhou2021mpi_compression} extend MVAPICH2 with both lossless and lossy compression for HPC workloads, while numerous frameworks~\cite{THC2024,compso2025,greedycompression2025,PHY2023,grace2021,li2024near_lossless,huang2023lowrank,takezawa2023compression,ZeRO2023,qsgd2017} adopt lossy compression for training and inference. While effective, these approaches introduce approximation error.
\section{Conclusion}
The rapid scaling of large language models has made GPU communication a key performance bottleneck. In this work, we introduced \sys{}, a unified approach that embeds lossless compression into GPU communication primitives. \sys{} supports both point-to-point and collective communication without sacrificing numerical fidelity. 
\sys{} accelerates RL weight synchronization by up to 47.5\% and reduces vLLM end-to-end inference latency by up to 10\%.

\bibliographystyle{ACM-Reference-Format}
\bibliography{ref}

@misc{NeuZip2024,
  author       = {Yongchang Hao and Yanshuai Cao and Lili Mou},
  title        = {NeuZip: Memory-Efficient Training and Inference with Dynamic Compression of Neural Networks},
  year         = {2024},
  eprint       = {2410.20650},
  archivePrefix= {arXiv},
  primaryClass = {cs.LG},
}

@inproceedings{ZipNN2025,
  author    = {Moshik Hershcovitch and Andrew Wood and Leshem Choshen and Guy Girmonsky and Roy Leibovitz and Ilias Ennmouri and Michal Malka and Peter Chin and Swaminathan Sundararaman and Danny Harnik},
  title     = {ZipNN: Lossless Compression for {AI} Models},
  booktitle = {2025 IEEE 18th International Conference on Cloud Computing (CLOUD)},
  year      = {2025},
  pages     = {186--198},
  doi       = {10.1109/CLOUD67622.2025.00028},
}

@misc{cui2025optimizingsloorientedllmserving,
  author       = {Cui, Weihao and Chen, Yukang and Zhao, Han and Xu, Ziyi and Chen, Quan and Chen, Xusheng and Yangjie, Zhou and Sun, Shixuan and Guo, Minyi},
  title        = {Optimizing {SLO}-oriented {LLM} Serving with {PD}-Multiplexing},
  year         = {2025},
  eprint       = {2504.14489v1},
  archivePrefix= {arXiv},
  primaryClass = {cs.OS},
}

@misc{msccl,
  author       = {Aashaka Shah and Abhinav Jangda and Binyang Li and Caio Rocha and Changho Hwang and Jithin Jose and Madan Musuvathi and Olli Saarikivi and Peng Cheng and Qinghua Zhou and Roshan Dathathri and Saeed Maleki and Ziyue Yang},
  title        = {{MSCCL++}: Rethinking {GPU} Communication Abstractions for Cutting-edge {AI} Applications},
  year         = {2025},
  eprint       = {2504.09014},
  archivePrefix= {arXiv},
  primaryClass = {cs.DC},
}

@inproceedings{megatron-lm,
  author    = {Narayanan, Deepak and Shoeybi, Mohammad and Casper, Jared and LeGresley, Patrick and Patwary, Mostofa and Korthikanti, Vijay and Vainbrand, Dmitri and Kashinkunti, Prethvi and Bernauer, Julie and Catanzaro, Bryan and Phanishayee, Amar and Zaharia, Matei},
  title     = {Efficient Large-Scale Language Model Training on {GPU} Clusters Using Megatron-{LM}},
  booktitle = {Proceedings of the International Conference for High Performance Computing, Networking, Storage and Analysis},
  series    = {SC '21},
  year      = {2021},
  articleno = {58},
  numpages  = {15},
  publisher = {Association for Computing Machinery},
  doi       = {10.1145/3458817.3476209},
}

@misc{deepseekai2025deepseekv3technicalreport,
  author       = {{DeepSeek-AI} and Aixin Liu and Bei Feng and others},
  title        = {{DeepSeek-V3} Technical Report},
  year         = {2025},
  eprint       = {2412.19437},
  archivePrefix= {arXiv},
  primaryClass = {cs.CL},
}

@article{google-palm,
  author    = {Chowdhery, Aakanksha and Narang, Sharan and Devlin, Jacob and Bosma, Maarten and Mishra, Gaurav and Roberts, Adam and Barham, Paul and Chung, Hyung Won and Sutton, Charles and Gehrmann, Sebastian and Schuh, Parker and Shi, Kensen and Tsvyashchenko, Sashank and Maynez, Joshua and Rao, Abhishek and Barnes, Parker and Tay, Yi and Shazeer, Noam and Prabhakaran, Vinodkumar and Reif, Emily and Du, Nan and Hutchinson, Ben and Pope, Reiner and Bradbury, James and Austin, Jacob and Isard, Michael and Gur-Ari, Guy and Yin, Pengcheng and Duke, Toju and Levskaya, Anselm and Ghemawat, Sanjay and Dev, Sunipa and Michalewski, Henryk and Garcia, Xavier and Misra, Vedant and Robinson, Kevin and Fedus, Liam and Zhou, Denny and Ippolito, Daphne and Luan, David and Lim, Hyeontaek and Zoph, Barret and Spiridonov, Alexander and Sepassi, Ryan and Dohan, David and Agrawal, Shivani and Omernick, Mark and Dai, Andrew M. and Pillai, Thanumalayan Sankaranarayana and Pellat, Marie and Lewkowycz, Aitor and Moreira, Erica and Child, Rewon and Polozov, Oleksandr and Lee, Katherine and Zhou, Zongwei and Wang, Xuezhi and Saeta, Brennan and Diaz, Mark and Firat, Orhan and Catasta, Michele and Wei, Jason and Meier-Hellstern, Kathy and Eck, Douglas and Dean, Jeff and Petrov, Slav and Fiedel, Noah},
  title     = {{PaLM}: Scaling Language Modeling with Pathways},
  journal   = {J. Mach. Learn. Res.},
  volume    = {24},
  number    = {1},
  year      = {2023},
  articleno = {240},
  numpages  = {113},
  publisher = {JMLR.org},
}

@misc{qwen2025qwen25technicalreport,
  author       = {Qwen and An Yang and Baosong Yang and others},
  title        = {{Qwen2.5} Technical Report},
  year         = {2025},
  eprint       = {2412.15115},
  archivePrefix= {arXiv},
  primaryClass = {cs.CL},
}

@inproceedings{colossalai,
  author    = {Li, Shenggui and Liu, Hongxin and Bian, Zhengda and Fang, Jiarui and Huang, Haichen and Liu, Yuliang and Wang, Boxiang and You, Yang},
  title     = {Colossal-{AI}: A Unified Deep Learning System For Large-Scale Parallel Training},
  booktitle = {Proceedings of the 52nd International Conference on Parallel Processing},
  series    = {ICPP '23},
  year      = {2023},
  pages     = {766--775},
  publisher = {Association for Computing Machinery},
  doi       = {10.1145/3605573.3605613},
}

@inproceedings{deepspeed,
  author    = {Jeff Rasley and Samyam Rajbhandari and Olatunji Ruwase and Yuxiong He},
  title     = {{DeepSpeed}: System Optimizations Enable Training Deep Learning Models with Over 100 Billion Parameters},
  booktitle = {Proceedings of the 26th ACM SIGKDD International Conference on Knowledge Discovery \& Data Mining},
  series    = {KDD '20},
  year      = {2020},
  pages     = {3505--3506},
  publisher = {Association for Computing Machinery},
  doi       = {10.1145/3394486.3406703},
}

@inproceedings{crux,
  author    = {Jiamin Cao and Yu Guan and Kun Qian and others},
  title     = {Crux: {GPU}-Efficient Communication Scheduling for Deep Learning Training},
  booktitle = {Proceedings of the ACM SIGCOMM 2024 Conference},
  series    = {ACM SIGCOMM '24},
  year      = {2024},
  pages     = {1--15},
  publisher = {Association for Computing Machinery},
  doi       = {10.1145/3651890.3672239},
}

@inproceedings{cassini,
  author    = {Sudarsanan Rajasekaran and Manya Ghobadi and Aditya Akella},
  title     = {{CASSINI}: Network-Aware Job Scheduling in Machine Learning Clusters},
  booktitle = {Proceedings of the 21st USENIX Symposium on Networked Systems Design and Implementation},
  series    = {NSDI '24},
  year      = {2024},
  articleno = {78},
  numpages  = {18},
  publisher = {USENIX Association},

}

@misc{chang2024flux,
  author       = {Li-Wen Chang and Wenlei Bao and Qi Hou and others},
  title        = {{FLUX}: Fast Software-based Communication Overlap On {GPUs} Through Kernel Fusion},
  year         = {2024},
  eprint       = {2406.06858},
  archivePrefix= {arXiv},
  primaryClass = {cs.LG},
}

@inproceedings{hu2025demystifying,
  author    = {Zhiyi Hu and Siyuan Shen and Tommaso Bonato and others},
  title     = {Demystifying {NCCL}: An In-Depth Analysis of {GPU} Communication Protocols and Algorithms},
  booktitle = {2025 IEEE Symposium on High-Performance Interconnects (HOTI)},
  year      = {2025},
  pages     = {48--59},
  doi       = {10.1109/HOTI66940.2025.00024},
}

@misc{nccl,
  author       = {Sylvain Jeaugey},
  title        = {{NCCL}: Optimized Primitives for Collective Multi-{GPU} Communication},
  year         = {2017},
  howpublished = {\url{https://developer.nvidia.com/nccl}},
}

@inproceedings{dfloat11,
  author    = {Tianyi Zhang and Mohsen Hariri and Shaochen Zhong and others},
  title     = {70\% Size, 100\% Accuracy: Lossless {LLM} Compression for Efficient {GPU} Inference via Dynamic-Length Float ({DF}loat11)},
  booktitle = {The Thirty-ninth Annual Conference on Neural Information Processing Systems},
  year      = {2025},
}

@misc{dietgpu_github,
  author       = {{Meta AI Research}},
  title        = {{DietGPU}},
  year         = {2026},
  howpublished = {\url{https://github.com/facebookresearch/dietgpu}},
  note         = {GitHub repository, accessed 2026-03-07},
}

@inproceedings{kwon2023efficient,
  author    = {Woosuk Kwon and Zhuohan Li and Siyuan Zhuang and others},
  title     = {Efficient Memory Management for Large Language Model Serving with {PagedAttention}},
  booktitle = {Proceedings of the ACM SIGOPS 29th Symposium on Operating Systems Principles},
  year      = {2023},
  doi       = {10.1145/3600006.3613165},
}

@inproceedings{azami2025lossless,
  author    = {Noushin Azami and Alex Fallin and Martin Burtscher},
  title     = {Efficient Lossless Compression of Scientific Floating-Point Data on {CPUs} and {GPUs}},
  booktitle = {Proceedings of the 30th ACM International Conference on Architectural Support for Programming Languages and Operating Systems, Volume 1},
  series    = {ASPLOS '25},
  year      = {2025},
  doi       = {10.1145/3669940.3707280},
}

@misc{nvcomp2023,
  author       = {{NVIDIA}},
  title        = {{nvCOMP}: {NVIDIA} {GPU} Data Compression Library},
  year         = {2023},
  howpublished = {\url{https://github.com/NVIDIA/nvcomp}},
  note         = {Accessed: July 31, 2023},
}

@misc{uccl_p2p,
  author       = {{UCCL Project}},
  title        = {{KV} Transfer Engine: High-Performance {GPU} Communication in {UCCL}},
  year         = {2024},
  howpublished = {\url{https://uccl-project.github.io/posts/kv-transfer-engine/}},
  note         = {Accessed: 2026},
}

@misc{rccl,
  author       = {{AMD}},
  title        = {{RCCL}: {AMD} {ROCm} Collective Communication Library},
  year         = {2024},
  howpublished = {\url{https://github.com/ROCmSoftwarePlatform/rccl}},
  note         = {Accessed: 2026},
}

@misc{mooncake_te,
  author       = {{Mooncake Project}},
  title        = {{Mooncake} Transfer Engine},
  year         = {2024},
  howpublished = {\url{https://github.com/kvcache-ai/Mooncake}},
  note         = {Accessed: 2026},
}

@misc{glm45v2025,
  author       = {{V Team} and Wenyi Hong and others},
  title        = {{GLM-4.5V} and {GLM-4.1V-Thinking}: Towards Versatile Multimodal Reasoning with Scalable Reinforcement Learning},
  year         = {2025},
  eprint       = {2507.01006},
  archivePrefix= {arXiv},
  primaryClass = {cs.CV},
}

@misc{qwen,
  author       = {Jinze Bai and Shuai Bai and Yunfei Chu and others},
  title        = {Qwen Technical Report},
  year         = {2023},
  eprint       = {2309.16609},
  archivePrefix= {arXiv},
  primaryClass = {cs.CL},
}

@inproceedings{MPC2015,
  author    = {Annie Yang and Hari Mukka and Farbod Hesaaraki and Martin Burtscher},
  title     = {{MPC}: A Massively Parallel Compression Algorithm for Scientific Data},
  booktitle = {2015 IEEE International Conference on Cluster Computing},
  year      = {2015},
  pages     = {381--389},
  doi       = {10.1109/CLUSTER.2015.59},
}

@misc{Han2015DeepCC,
  author       = {Song Han and Huizi Mao and William J. Dally},
  title        = {Deep Compression: Compressing Deep Neural Network with Pruning, Trained Quantization and Huffman Coding},
  year         = {2015},
  eprint       = {1510.00149},
  archivePrefix= {arXiv},
  primaryClass = {cs.CV},
}

@misc{TDT2025,
  author       = {Samirasadat Jamalidinan and Kazem Cheshmi},
  title        = {Floating-Point Data Transformation for Lossless Compression},
  year         = {2025},
  eprint       = {2506.18062},
  archivePrefix= {arXiv},
  primaryClass = {cs.DB},
}

@inproceedings{gzcccl2024,
  author    = {Jiajun Huang and Sheng Di and Xiaodong Yu and others},
  title     = {g{ZCCL}: Compression-Accelerated Collective Communication Framework for {GPU} Clusters},
  booktitle = {Proceedings of the 38th ACM International Conference on Supercomputing},
  series    = {ICS '24},
  year      = {2024},
  pages     = {437--448},
  doi       = {10.1145/3650200.3656636},
}

@inproceedings{ghzccl2025,
  author    = {Jiajun Huang and Sheng Di and Yafan Huang and others},
  title     = {{GhZCCL}: Advancing {GPU}-aware Collective Communications with Homomorphic Compression},
  booktitle = {Proceedings of the 39th ACM International Conference on Supercomputing},
  series    = {ICS '25},
  year      = {2025},
  pages     = {43--56},
  doi       = {10.1145/3721145.3733642},
}

@misc{zccl2025,
  author       = {Jiajun Huang and Sheng Di and Xiaodong Yu and others},
  title        = {{ZCCL}: Significantly Improving Collective Communication With Error-Bounded Lossy Compression},
  year         = {2025},
  eprint       = {2502.18554},
  archivePrefix= {arXiv},
  primaryClass = {cs.DC},
}

@inproceedings{huang2024mpi,
  author    = {Jiajun Huang and Sheng Di and Xiaodong Yu and others},
  title     = {An Optimized Error-Controlled {MPI} Collective Framework Integrated with Lossy Compression},
  booktitle = {2024 IEEE International Parallel and Distributed Processing Symposium (IPDPS)},
  year      = {2024},
  pages     = {752--764},
}

@inproceedings{zhou2021mpi_compression,
  author    = {Q. Zhou and C. Chu and N. S. Kumar and others},
  title     = {Designing High-Performance {MPI} Libraries with On-the-fly Compression for Modern {GPU} Clusters},
  booktitle = {2021 IEEE International Parallel and Distributed Processing Symposium (IPDPS)},
  series    = {IPDPS '21},
  year      = {2021},
  pages     = {444--453},
  doi       = {10.1109/IPDPS49936.2021.00053},
}

@misc{THC2024,
  author       = {Minghao Li and Ran Ben Basat and Shay Vargaftik and others},
  title        = {{THC}: Accelerating Distributed Deep Learning Using Tensor Homomorphic Compression},
  year         = {2024},
  eprint       = {2302.08545},
  archivePrefix= {arXiv},
  primaryClass = {cs.LG},
}

@inproceedings{compso2025,
  author    = {Baixi Sun and Weijin Liu and J. Gregory Pauloski and others},
  title     = {{COMPSO}: Optimizing Gradient Compression for Distributed Training with Second-Order Optimizers},
  booktitle = {Proceedings of the 30th ACM SIGPLAN Symposium on Principles and Practice of Parallel Programming},
  series    = {PPoPP '25},
  year      = {2025},
  pages     = {212--224},
  doi       = {10.1145/3710848.3710852},
}

@misc{greedycompression2025,
  author       = {Chuyan Chen and Yutong He and Pengrui Li and others},
  title        = {Greedy Low-Rank Gradient Compression for Distributed Learning with Convergence Guarantees},
  year         = {2025},
  eprint       = {2507.08784},
  archivePrefix= {arXiv},
  primaryClass = {cs.LG},
}

@article{PHY2023,
  author  = {Chen-Chun Chen and Yu-Min Chou and Jerry Chou},
  title   = {{PHY}: A Performance-Driven Hybrid Communication Compression Method for Distributed Training},
  journal = {Journal of Parallel and Distributed Computing},
  volume  = {180},
  pages   = {104719},
  year    = {2023},
  doi     = {10.1016/j.jpdc.2023.104719},
}

@inproceedings{li2024near_lossless,
  author    = {Xue Li and Cheng Guo and Kun Qian and others},
  title     = {Near-Lossless Gradient Compression for Data-Parallel Distributed {DNN} Training},
  booktitle = {Proceedings of the ACM Symposium on Cloud Computing},
  series    = {SoCC '24},
  year      = {2024},
  pages     = {977--994},
  doi       = {10.1145/3698038.3698541},
}

@inproceedings{grace2021,
  author    = {Hang Xu and Chen-Yu Ho and Ahmed M. Abdelmoniem and others},
  title     = {{GRACE}: A Compressed Communication Framework for Distributed Machine Learning},
  booktitle = {2021 IEEE 41st International Conference on Distributed Computing Systems (ICDCS)},
  series    = {ICDCS '21},
  year      = {2021},
  pages     = {561--572},
  doi       = {10.1109/ICDCS51616.2021.00060},
}

@article{huang2023lowrank,
  author    = {Siyuan Huang and Brian D. Hoskins and Matthew W. Daniels and others},
  title     = {Low-Rank Gradient Descent for Memory-Efficient Training of Deep In-Memory Arrays},
  journal   = {Journal of Emerging Technologies in Computing Systems},
  year      = {2023},
  volume    = {19},
  number    = {2},
  articleno = {16},
  numpages  = {24},
  doi       = {10.1145/3577214},
}

@article{takezawa2023compression,
  author  = {Yuki Takezawa and Kenta Niwa and Makoto Yamada},
  title   = {Communication Compression for Decentralized Learning With Operator Splitting Methods},
  journal = {IEEE Transactions on Signal and Information Processing over Networks},
  year    = {2023},
  volume  = {9},
  pages   = {581--595},
  doi     = {10.1109/TSIPN.2023.3307894},
}

@misc{ZeRO2023,
  author       = {Guanhua Wang and Heyang Qin and Sam Ade Jacobs and others},
  title        = {{ZeRO++}: Extremely Efficient Collective Communication for Giant Model Training},
  year         = {2023},
  eprint       = {2306.10209},
  archivePrefix= {arXiv},
  primaryClass = {cs.DC}
}

@inproceedings{qsgd2017,
  author    = {Dan Alistarh and Demjan Grubic and Jerry Li and others},
  title     = {{QSGD}: Communication-Efficient {SGD} via Gradient Quantization and Encoding},
  booktitle = {Advances in Neural Information Processing Systems},
  series    = {NeurIPS '17},
  year      = {2017},
}

@inproceedings{fei2021sparse,
  author    = {Jiawei Fei and Chen-Yu Ho and Atal N. Sahu and others},
  title     = {Efficient Sparse Collective Communication and Its Application to Accelerate Distributed Deep Learning},
  booktitle = {Proceedings of the 2021 ACM SIGCOMM Conference},
  series    = {SIGCOMM '21},
  year      = {2021},
  pages     = {676--691},
  doi       = {10.1145/3452296.3472904},
}

@misc{nixl2025,
  author       = {{NVIDIA}},
  title        = {{NIXL}: {NVIDIA} Inference Xfer Library},
  year         = {2025},
  howpublished = {\url{https://github.com/ai-dynamo/nixl}},
}

@inproceedings{langer2021nvshmem,
  author    = {Alexander Langer and Samuel Howell and Sreeram Potluri and others},
  title     = {Dynamic Symmetric Heap Allocation in {NVSHMEM}},
  booktitle = {OpenSHMEM and Related Technologies},
  series    = {Lecture Notes in Computer Science},
  publisher = {Springer},
  year      = {2021},
  pages     = {187--198},
  doi       = {10.1007/978-3-031-04888-3_12},
}

@misc{pplx2025,
  author       = {Nandor Licker and Kevin Hu and Vladimir Zaytsev and Lequn Chen},
  title        = {{RDMA} Point-to-Point Communication for {LLM} Systems},
  year         = {2025},
  eprint       = {2510.27656},
  archivePrefix= {arXiv},
  primaryClass = {cs.DC},
}

@misc{uccl,
  author       = {Yang Zhou and Zhongjie Chen and Ziming Mao and others},
  title        = {An Extensible Software Transport Layer for {GPU} Networking},
  year         = {2025},
  eprint       = {2504.17307},
  archivePrefix= {arXiv},
  primaryClass = {cs.NI},
}

@misc{grattafiori2024llama3,
  author       = {Aaron Grattafiori and Abhimanyu Dubey and Abhinav Jauhri and others},
  title        = {The {Llama} 3 Herd of Models},
  year         = {2024},
  eprint       = {2407.21783},
  archivePrefix= {arXiv},
  primaryClass = {cs.AI},
}

@misc{nvidiaCUDAprogramming,
  author       = {{NVIDIA}},
  title        = {{NVIDIA} {CUDA} {C} Programming Guide},
  year         = {2023},
  howpublished = {\url{https://docs.nvidia.com/cuda/cuda-c-programming-guide/}},
}

@inproceedings{fan2026zipserv,
  author    = {Ruibo Fan and Xiangrui Yu and Xinglin Pan and Zeyu Li and Weile Luo and Qiang Wang and Wei Wang and Xiaowen Chu},
  title     = {{ZipServ}: Fast and Memory-Efficient {LLM} Inference with Hardware-Aware Lossless Compression},
  booktitle = {Proceedings of the 31st ACM International Conference on Architectural Support for Programming Languages and Operating Systems, Volume 2},
  series    = {ASPLOS '26},
  year      = {2026},
  pages     = {2264--2280},
  publisher = {Association for Computing Machinery},
  address   = {New York, NY, USA},
  doi       = {10.1145/3779212.3790250},
}

@misc{he2025nondeterminism,
  author       = {Horace He and {Thinking Machines Lab}},
  title        = {Defeating Nondeterminism in {LLM} Inference},
  year         = {2025},
  howpublished = {\url{https://thinkingmachines.ai/blog/defeating-nondeterminism-in-llm-inference/}},
  note         = {Thinking Machines Lab: Connectionism},
}

@inproceedings{xu2025llm265,
  author    = {Ceyu Xu and Yongji Wu and Xinyu Yang and Beidi Chen and Matthew Lentz and Danyang Zhuo and Lisa Wu Wills},
  title     = {{LLM.265}: Video Codecs are Secretly Tensor Codecs},
  booktitle = {Proceedings of the 58th IEEE/ACM International Symposium on Microarchitecture},
  series    = {MICRO '25},
  year      = {2025},
  pages     = {445--460},
  publisher = {Association for Computing Machinery},
  address   = {New York, NY, USA},
  doi       = {10.1145/3725843.3756078},
}

@inproceedings{gauss,
author = {Gao, Tianxiang and Huo, Xiaokai and Liu, Hailiang and Gao, Hongyang},
title = {Wide neural networks as Gaussian processes: lessons from deep equilibrium models},
year = {2023},
publisher = {Curran Associates Inc.},
address = {Red Hook, NY, USA},
booktitle = {Proceedings of the 37th International Conference on Neural Information Processing Systems},
articleno = {2397},
numpages = {34},
location = {New Orleans, LA, USA},
series = {NIPS '23}
}

\appendix

\end{document}